\documentclass[a4paper,11pt]{article}
\usepackage[text={16.8cm,22.4cm}]{geometry}
\usepackage{amsmath,amsfonts,slashed,amssymb,tikz,bm,psfrag,graphicx,color,dsfont}
\usepackage{multicol,multirow}
\usepackage{float}
\usepackage{authblk}
\usepackage[nottoc]{tocbibind}
\usepackage[english]{babel}
\usepackage{array} 
\usepackage{bbm}

\usepackage{graphicx}
\usepackage{dcolumn}
\usepackage{bm}
\usepackage{xcolor}
\usepackage{colortbl,booktabs}
\usepackage{float}
\usepackage{slashed}
\usepackage{multirow}
\usepackage{multicol}
\usepackage{subfigure}
\usepackage[citecolor=blue]{hyperref} 
\allowdisplaybreaks 
\usepackage[figuresright]{rotating}
\usepackage{cancel}
\RequirePackage[sort&compress,square,comma,numbers]{natbib}
\allowdisplaybreaks
\title{\bf Revisiting $B_s \to PP$ and $PV$ Decays with Contributions from ${\phi}_{B2}$ with perturbativen QCD approach}

\vspace{0.5cm}
\author[]{Yueling Yang$^1$}
\author[]{Zhao-Jie L\"{u}$^1$}
\author[]{Su-Ping Jin$^{1}$
\thanks{E-mail: jinsuping@htu.edu.cn (corresponding author)}}
\author[]{Junfeng Sun$^1$}

\vspace{0.7cm}
\affil[]{$^1$Institute of Particle and Nuclear Physics, Henan Normal University, Xinxiang 453007, China}

\begin{document}

\maketitle

\medskip

\vspace{0.2cm}
\begin{abstract}

The $B_s \to PP$ and $PV$ decay modes are revisited at leading order within the perturbative QCD approach, incorporating the $B_s$ mesonic wave function (WF) ${\phi}_{B2}$. Here, $P$ represents the pseudoscalar mesons ${\pi}$ and $K$, while $V$ denotes the ground-state vector mesons.  The investigation incorporates two key refinements: the contribution of the sub-leading twist WF  $\phi_{B2}$ of the $B_s$ meson and the effects of higher-order terms in the distribution amplitudes (DAs) of the final-state mesons. Employing the minimum $\chi^2$ method, we optimize the shape parameter $\omega_{B_s}$ of the $B_s$ meson WF and systematically calculate the branching ratios and $CP$ violation parameters for these decay modes. Our results demonstrate that the inclusion of $\phi_{B2}$ significantly impacts both the branching ratios and $CP$ asymmetries, offer an improved agreement with existing experimental data for specific channels. This underscores the necessity of accounting for $\phi_{B2}$ in theoretical studies of $B_s$ weak decays. While the higher-order corrections in the final-state meson DAs yield comparatively smaller effects, they still enhance the theoretical predictions. These findings highlight the importance of refining both wave function modeling and higher-order contributions in pQCD calculations. Future high-precision experimental measurements will further test these predictions, while continued theoretical efforts are essential to explore additional interaction mechanisms and systematic uncertainties. The interplay between experimental advancements and theoretical improvements remains critical for a deeper understanding of $B_s$ meson decay dynamics.

\end{abstract}
\vfil

\newpage


\newpage

\section{Introduction}

$B$ meson physics represents a vital frontier in particle physics, driven by continuous advancements in both experimental techniques and theoretical understanding. The field is entering an era of unprecedented data collection through major experimental initiatives: the Belle-II experiment at the SuperKEKB $ e^+e^-$  collider is progressing toward its design goal of 50  $\text{ab}^{-1}$ integrated luminosity \cite{Belle-II:2018jsg}, while the upgraded LHCb detector at the High Luminosity LHC(HL-LHC) anticipates collecting approximately  300  $\text{fb}^{-1}$ \cite{LHCb:2018roe}. These efforts will be complemented by next-generation $e^+ e^-$ collider projects – notably the Circular Electron-Positron Collider (CEPC) \cite{CEPCStudyGroup:2023quu} and FCC-ee \cite{FCC:2018byv}– which are projected to generate over 
$10^{10}$ $B_s$  mesons via $Z^0$  boson decays. This impressive yield arises from two key factors: the substantial $Z^0 \to b\bar{b}$  branching fraction of  $\mathcal{B}(Z^0 \to b\bar{b}) \approx 15.12\%$  \cite{ParticleDataGroup:2024cfk} , combined with the b-quark fragmentation fraction to $B_s$ mesons of $f(b \to B_s) \approx 10.3\%$ \cite{DELPHI:2003pao,LHCb:2021qbv}. The relatively low fragmentation fraction compared to $B_{u,d}$  mesons ($\sim 40\%$)  reflects the suppressed production mechanism for $B_s$ during hadronization \cite{LHCb:2021qbv}, making high-statistics studies particularly valuable for probing rare processes.  

The particular importance of $B_s \to PP$ and $PV$ decays (where $P$ denotes light pseudoscalar mesons and $V$ represents light vector mesons) lies in their unique potential to probe both Standard Model predictions and potential New Physics scenarios. These processes serve as sensitive testing grounds for understanding QCD dynamics in heavy quark decays, investigating CP violation mechanisms, and constraining the unitarity triangle parameters. Moreover, they offer complementary information to their $B_d$ and $B_u$ counterparts due to the distinct penguin contributions and oscillation characteristics of the $B_s$ system. The interplay between theoretical calculations and experimental results in this sector continues to be crucial for verifying the consistency of the SM and identifying possible anomalies that might indicate physics beyond current paradigms.

Over the past years, significant progress has been made in understanding two-body charmless hadronic decays $B_s \to PP$ and $PV$, with extensive experimental measurements and theoretical investigations.  Among the possible $B_s \to PP$ and $PV$ decay channels, experimental collaborations including CDF \cite{CDF:2011who,CDF:2008llm,CDF:2011ubb,CDF:2014pzb}
, Belle \cite{Belle:2010yix,Belle:2015gho}, and LHCb \cite{LHCb:2016inp,LHCb:2012ihl,LHCb:2020wrt,LHCb:2019vww,LHCb:2014lcy,LHCb:2018pff,LHCb:2020byh,LHCb:2013clb} have successfully observed  seven channels to date. The current experimental status, including measurements of branching ratios and $CP$-violating asymmetries for these decays, has been systematically compiled in Table\ref{tab:bracpexp}. With the anticipated high-precision data from upgraded facilities like LHCb and Belle-II
\cite{ParticleDataGroup:2024cfk,HeavyFlavorAveragingGroupHFLAV:2024ctg}, significant improvements in both statistical significance and systematic uncertainties are expected in forthcoming measurements.
From the theoretical perspective, $B_s \to PP$ and $PV$  decays have been extensively investigated through various approaches, including QCD factorization \cite{Beneke:2003zv,Sun:2002rn,Cheng:2009mu,Chang:2014yma}, soft-collinear effective theory (SCET) \cite{Williamson:2006hb}, and perturbative QCD (pQCD) factorization  approach
\cite{Li:2004ep,Ali:2007ff,Xiao:2011tx,Wang:2014mua,Yan:2019nhf,Yan:2017nlj}, as well as flavor SU(3) symmetry analyses \cite{Cheng:2011qh,Cheng:2014rfa}. While these methods exhibit substantial differences in their predictions for CP-violating asymmetry patterns and magnitudes, they demonstrate reasonable consistency in branching ratio predictions when considering current theoretical uncertainties.

\begin{table}
\caption{ The measured values of the branching ratios (in units of $10^{-6}$) of the seven considered decay modes and $A_{\rm CP}(B^0_s\to \pi^+ K^-)$ of $B^0_s\to K^+ K^-$  decay,
as reported by the Belle\cite{Belle:2010yix,Belle:2015gho}, CDF\cite{CDF:2011who,CDF:2008llm,CDF:2011ubb,CDF:2014pzb},
LHCb Collaboration \cite{LHCb:2016inp,LHCb:2012ihl,LHCb:2020wrt,LHCb:2019vww,LHCb:2014lcy,LHCb:2018pff,LHCb:2020byh,LHCb:2013clb}, and the world averages as given in Refs.~\cite{HeavyFlavorAveragingGroupHFLAV:2024ctg,ParticleDataGroup:2024cfk}.}

\label{tab:bracpexp}
\begin{tabular*}{16cm}{@{\extracolsep{\fill}}l|ccc|cc}  \hline\hline
\multicolumn{1}{c|}{\rm Mode} &{\rm Belle}\cite{Belle:2010yix,Belle:2015gho} &{\rm CDF}\cite{CDF:2011who,CDF:2008llm,CDF:2011ubb,CDF:2014pzb}&{\rm LHCb}\cite{LHCb:2016inp,LHCb:2012ihl,LHCb:2020wrt,LHCb:2019vww,LHCb:2014lcy,LHCb:2018pff,LHCb:2020byh,LHCb:2013clb}&{\rm PDG}\cite{ParticleDataGroup:2024cfk}&{\rm HFLAV}\cite{HeavyFlavorAveragingGroupHFLAV:2024ctg} \\ \hline
${\cal B}(\bar B_s^0\to \pi^- K^+)$&      $<26$&     $5.8\pm1.12    $&$6.0\pm0.7\pm0.6$&$5.9\pm0.7 $ &$6.1^{+0.9}_{-0.8}$\\
${\cal B}(\bar B_s^0\to \pi^+\pi^-)$&      $<12$&     $0.65\pm0.19   $&$0.75\pm0.09\pm0.07$ &$0.72\pm0.1 $ &$0.74^{+0.12}_{-0.10}$\\
${\cal B}(\bar B_s^0\to K^+ K^-)$&      $38^{+10}_{-9}\pm7$&     $28.3\pm3.6    $&$25.7\pm1.7\pm2.5$ &$27.2\pm2.3  $&$27.4^{+3.2}_{-2.8} $\\
${\cal B}(\bar B_s^0\to K^0 \bar K^0)$&      $19.6^{+5.8}_{-5.1}\pm2.0$&     $-$&$1.68\pm3.4^{+1.6}_{-1.5}$ &$17.6\pm3.1$ &$17.4\pm 3.1 $\\
\hline
${\cal B}(\bar B_s^0\to K^\pm \bar K^{*\mp})$&      $-$&     $-$&$18.6\pm1.2\pm4.5$ &$19.0\pm5$ &$18.6\pm 4.7 $\\
${\cal B}(\bar B_s^0\to  \bar K^{*0} K^0)$&      $-$&     $-$&$19.8\pm2.8\pm5$ &$20.0\pm6.0$ &$19.8\pm 5.7 $\\
${\cal B}(\bar B_s^0\to  \pi^- K^{*+})$&      $-$&     $-$&$2.9\pm1.0\pm0.2$ &$2.9\pm1.1$ &$3.0\pm 1.1 $\\
\hline
$A_{\rm CP}( B_s^0\to K^-\pi^+)$         &      $0.22\pm 0.07$   &$-$ &    $0.236\pm0.017  $                              &$0.224\pm0.012$                      & $-$ \\
 &                                                            & $-$  &  $0.213\pm0.017 $                                                &                                                         &     \\ 
 \hline\hline
\end{tabular*}
\end{table}

In our previous investigation \cite{Yang:2020xal,Yang:2022ebu}, we conducted a systematic reanalysis of $B \to PP$ and $PV$ decays within the pQCD framework. These studies incorporated two crucial enhancements: the inclusion of subleading-twist wave functions (WFs) for the $B$ meson and updated distribution amplitudes (DAs) for final-state pseudoscalar mesons. Notably, our findings revealed that the $B$-mesonic WF component ${\phi}_{B2}$ - frequently overlooked in prior studies - exerts distinct influence on both hadronic matrix elements (HMEs) and branching ratios, with magnitude comparable to next-to-leading-order (NLO) corrections.

Building upon these insights, the present work extends this analysis to charmless $B_s \to PP$ and $PV$ decay channels. This extension aligns with recent advancements in both theoretical precision and experimental resolution. Our current investigation maintains focus on $P = \{\pi, K\}$ final states, deliberately excluding $\eta$ and $\eta^\prime$ mesons due to persistent theoretical challenges in resolving their flavor mixing structure and potential glueball admixtures. By incorporating this subleading-twist WFs for the $B_s$ meson, we systematically enhance the theoretical framework, leading to the following findings:
\begin{itemize}
\item{We demonstrate that the $\phi_{B_2}$ term contributes zero in factorizable annihilation topologies. This property provides a crucial basis for simplifying theoretical calculations of the corresponding decay channels.}

\item{By fully incorporating the $\phi_{B_2}$ term, we observe substantial corrections to the branching ratios of various decay channels, ranging from $1.4\%$ to $59.4\%$. Notably, in the decay channel $\bar{B}_s^0 \rightarrow K^{-} K^{+}$, this correction reduces the discrepancy between theoretical predictions and the latest LHCb experimental data from $43\%$ to $17\%$, thereby validating the necessity of higher-order wave function terms for enhancing theoretical precision. In certain decay channels, such as $\bar{B}_s^0 \rightarrow \pi^{0} K^{0}$, $\bar{B}_s^0 \rightarrow K^{-} K^{+}$, and $\bar{B}_s^0 \rightarrow \bar{K}^{0} K^0$, the contribution of the $\phi_{B_2}$ term is comparable to that of next-to-leading order corrections.}

\item{Numerical results indicate that variations in the Gegenbauer moments $a_2^{P(V)}$ within the specified input parameter ranges lead to average fluctuations of approximately $7\%$ to $10\%$ in the branching ratios. This underscores the importance of accurately optimizing non-perturbative parameters.}

\item{After including traditionally neglected higher-order terms in the DAs of final-state mesons, we find that the changes in branching ratios are less than $6\%$. This result demonstrates the stability of the existing theoretical framework concerning higher-order terms and supports the validity of certain approximations within the pQCD approach.}

\end{itemize}

The structure of this paper is organized as follows. In Section \ref{sec:hamiltonian}, we briefly outline the theoretical framework underlying the analysis including effective Hamiltonian and decay amplitudes. Section \ref{sec:kinematics} defines the relevant kinematic and provides the expressions for WFs involved in the decays. We reevaluate the branching ratios and $CP$ asymmetries for the $B_s \to PP$ and $PV$ decays, considering the effects of the $B_s$-mesonic wave function $\phi_{B2}$ and present the numerical results in Section \ref{sec:numeric}. The paper concludes with a summary in Section \ref{sec:summary}. Additional details on the building blocks and the decay amplitudes for the $B_s \to PP$ and $PV$ decays can be found in Appendix \ref{sec:mode}.

\section{EFFECTIVE HAMILTONIAN AND DECAY AMPLITUDES \label{sec:hamiltonian}}

At the quark level, The low-energy effective Hamiltonian for the decays $\bar{B}^0_s \rightarrow PP$ and $PV$ can be expressed as the product of Wilson coefficients and effective operators,

\begin{align}
 \mathcal{H}_{\mathrm{eff}}=\frac{G_{F}}{\sqrt{2}}\sum\limits_{q=d,s}\Big\{V_{ub}V_{uq}^{*}
  \sum\limits_{i=1}^{2}C_{i}(\mu)\mathcal{O}_{i}-V_{tb}V_{tq}^{*}
  \sum\limits_{j=3}^{10}C_{j}(\mu)\mathcal{O}_{j}\Big\}+h.c.,
\end{align}
where $G_F$ is the Fermi coupling constant, $V_i$ are the elements of the Cabibbo-Kobayashi-Maskawa (CKM) matrix, $C_i(\mu)$ are the Wilson coefficients evaluated at the energy scale $\mu$, and $\mathcal{O}_i$ are the effective operators corresponding to specific decay processes. 

Utilizing the phenomenological Wolfenstein parametrization up to $\mathcal{O}(\lambda^8)$, the CKM factors read
\begin{align}
  V_{ub}\,V_{us}^{\ast} &=
    A\,{\lambda}^{4}\,({\rho}-i\,{\eta})
    +{\cal O}({\lambda}^{8})
    \label{ckm-vub-vus}, \\
    V_{ub}\,V_{ud}^{\ast} &=
    A\,{\lambda}^{3}\,({\rho}-i\,{\eta})\,
    (1-\frac{1}{2}\,{\lambda}^{2}-\frac{1}{8}\,{\lambda}^{4})
    +{\cal O}({\lambda}^{8})
    \label{ckm-vub-vud}, \\
    V_{tb}\,V_{td}^{\ast} &=
    A\,{\lambda}^{3}
    +A^{3}\,{\lambda}^{7}\,({\rho}-i\,{\eta}-\frac{1}{2})
    -V_{ub}\,V_{ud}^{\ast}
    +{\cal O}({\lambda}^{8})
    \label{ckm-vtb-vtd}, \\
    V_{tb}\,V_{ts}^{\ast} &=
    -A\,{\lambda}^{2}\,(1-\frac{1}{2}\,{\lambda}^{2}-\frac{1}{8}\,{\lambda}^{4})
    +\frac{1}{2}\,A^{3}\,{\lambda}^{6}
    -V_{ub}\,V_{us}^{\ast}
    +{\cal O}({\lambda}^{8})
    \label{ckm-vtb-vts}.
    \end{align}
The parameters $ A $, $ \lambda $, $ \rho $, and $ \eta $ correspond to the Wolfenstein parametrization of the CKM matrix. The most recent fitted values for these parameters can be found in Ref.\cite{ParticleDataGroup:2024cfk}. A summary of these fitted values is provided in Table \ref{tab:input}.

In the SM, the decay amplitude for $ \bar{B}^0_s \to PM $ ($M=P,V$) can be expressed as:
\begin{equation}
    \mathcal{A}(\bar{B}^0_s \to PM) = \langle PM | \mathcal{H}_{\text{eff}} | \bar{B}^0_s \rangle = \frac{G_F}{\sqrt{2}} \sum_{i=1}^{10} V_{i}\, C_{i} \langle PM | \mathcal{O}_i | \bar{B}^0_s \rangle.
\end{equation}
The HMEs $ \langle \mathcal{O}_i \rangle \equiv \langle PM | \mathcal{O}_i | \bar{B}^0_s \rangle $ describe the transition from quarks to hadrons. These matrix elements involve energy scales that lie between the perturbative and non-perturbative regimes. In the pQCD approach, by retaining the transverse momentum of the quarks and introducing Sudakov factors, the HMEs can be written in a convolution form, incorporating both the perturbative (hard scattering kernel) and non-perturbative (universal hadronic WF) parts. Thus, the decay amplitude can be factorized into the convolution of three parts: the Wilson coefficients $ C_i $, the hard scattering kernel $ H_i $, and the hadronic WFs.
Therefore, the HMEs for the $ \bar{B}^0_s \to PM $ decay process can be written as

\begin{align}
    \langle PM | C_i \, \mathcal{O}_i | \bar{B}^0_s \rangle &\propto \int dx_1 \, dx_2 \, dx_3 \, db_1 \, db_2 \, db_3 \, C_i(t_i) \mathcal{H}_i(x_1, x_2, x_3, b_1, b_2, b_3) \nonumber \\
    &\qquad\times\Phi_{B_s}(x_1, b_1) e^{-S_{B_s}} \, \Phi_{P}(x_2, b_2) e^{-S_{P}} \, \Phi_{M}(x_3, b_3) e^{-S_{M}},
\end{align}
where $ b_i $ denotes the conjugate variable of the transverse momentum $ \vec{k}_{i{\perp}} $ of the valence quarks; $ \mathcal{H}_i $ represents the scattering amplitudes for hard gluon exchange interactions among the quarks; and $ e^{-S_i} $ is the Sudakov factor. Additional variables and input parameters are described in the following section.

\section{KINEMATICS AND THEWAVE FUNCTIONS \label{sec:kinematics}}
\begin{figure}[htbp]
\begin{minipage}[t]{1\linewidth}
  \centering
  \includegraphics[width=0.6\columnwidth]{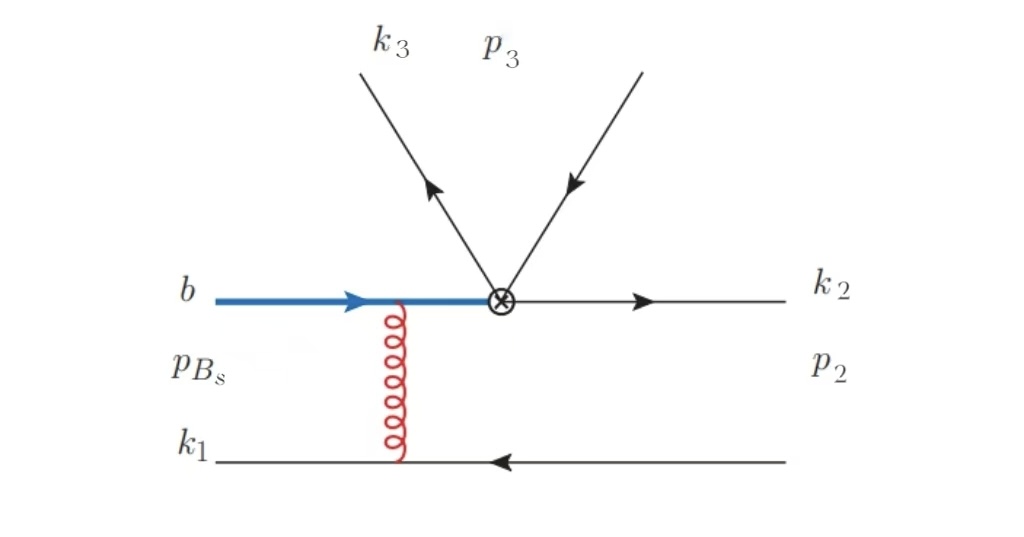}
\end{minipage}
  \caption{A LO diagram for the $\bar B_s(p_{B_s})\to M_2(p_2)M_3(p_3)$ decay.}\label{fig:momentumfractions}
\end{figure}

The momenta of the $\bar{B}^0_s$ meson $p_{B_s}$, recoiling meson $M_2$ ($p_2$) , and emitted meson $M_3$ ($p_3$), along with their corresponding parton momenta, are defined in light-cone coordinates as 
\begin{eqnarray}
	p_B = \frac{m_B}{\sqrt{2}}(1, 1, 0), \ \ \ \ \ \ k_1 = (x_1 \frac{m_B}{\sqrt{2}}, 0, \vec{k}_{1\perp}), \nonumber \\
	p_2 =  \frac{m_B}{\sqrt{2}}(0, 1, 0), \ \ \ \ \ \ k_2 = (0, x_2 \frac{m_B}{\sqrt{2}},  \vec {k}_{2\perp}),  \\
	p_3 =  \frac{m_B}{\sqrt{2}}(1, 0, 0), \ \ \ \ \ \ k_3 = (x_3 \frac{m_B}{\sqrt{2}}, 0, \vec{k}_{3\perp}), \nonumber \label{eq:momentum}
\end{eqnarray}
which are shown in Fig.~\ref{fig:momentumfractions}. Here $m_{B_s}$ denotes the $B_s$ meson mass, while
 $x_i$ represent the longitudinal momentum fractions carried by the constituent partons.
The transverse momentum components $\Vec{k}_{i\perp}$ in Eq~.(\ref{eq:momentum}) are treated as suppressed degrees of freedom.
For the vector meson, we also take into account its longitudinal polarization vector, given by $e^{\|}_V=\displaystyle \frac{p_V}{m_V}-\frac{m_V}{p_V \cdot n_{-}}n_{-}$.

According to the conventions established in the referenced works \cite{Ali:2007ff,Ball:2006wn,Yang:2020xal}, $\bar{B}^0_s$ meson WFs can be expressed as
\begin{eqnarray} 
  {\langle}0{\vert} \bar{s}_{\alpha}(z)~b_{\beta}(0){\vert}
   \bar B_s^0(p){\rangle}=\frac{f_{B_s}}{4N_c}{\int} {d}^{4}k~e^{-i\,k_{1}{\cdot}z}\,
   \Big\{ \big(\!\!\not{p}+m_{B_s}\big)~ {\gamma}_{5}~
   \Big[~\frac{\not{n}_{-}}{\sqrt{2}}~{\phi}_{B}^{+}
  +\frac{\not{n}_{+}}{\sqrt{2}}~{\phi}_{B}^{-} \Big] \Big\}_{{\beta}{\alpha}}.
   \label{B-meson-WF-definition}
   \end{eqnarray}
In the above equation, $ f_{B_s} $ denotes the decay constant, $ n_{+} = (1,0,0) $ represents the positive direction, and $ n_{-} = (0,1,0) $ corresponds to the negative direction. The scalar functions ${\phi}_{B}^{+}$ and ${\phi}_{B}^{-}$ correspond to the leading and subleading twist WFs, respectively.
These wave functions exhibit distinct asymptotic behaviors as the longitudinal momentum fraction of the light quark $x_{1}$ ${\to}$ $0$. Their relationship is given by the equation
    \begin{equation}
   {\phi}_{B}^{+}(x_{1}) + x_{1}\, {\phi}_{B}^{-{\prime}}(x_{1})\, =\, 0
    \label{wf-b-meson-equation-of-motion},
    \end{equation}
which reflects the equation of motion for the wave functions. We  can define 
${\phi}_{B1}$ and  ${\phi}_{B2}$ based on the following relations
   \begin{align}
       {\phi}_{B1}\, =\, {\phi}_{B}^{+},\qquad 
       {\phi}_{B2}\, =\, {\phi}_{B}^{+}-{\phi}_{B}^{-}.
   \end{align}

\begin{figure}[H]
\centering
 ~~~~\includegraphics[width=8cm]{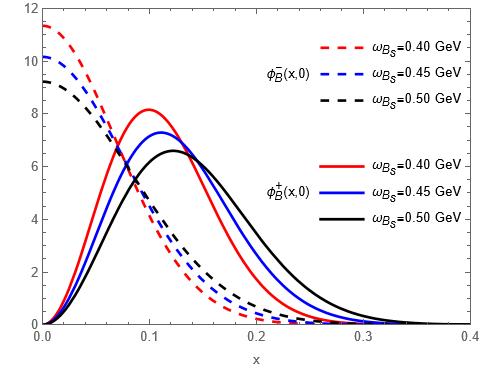}
\caption{The shapes of the  $B_s$ DAs (${\phi}_{B}^{+}$ and ${\phi}_{B}^{-}$) as a function of  the longitudinal momentum fraction $x$ (horizontal axis)}
  \label{fig-wf-1}
\end{figure}

Although the expressions for $\phi_B^+$ and $\phi_B^-$ generally differ, the equation of motion (Eq. \ref{wf-b-meson-equation-of-motion}) often leads to the approximation $\phi_B^+ = \phi_B^-$ in many phenomenological studies of nonleptonic $B$ meson decays. In this approximation, only the contributions from $\phi_{B1}$ are considered, and those from $\phi_{B2}$ are neglected. However, several studies \cite{Huang:2004hw,Yang:2020xal,Yang:2022ebu,Kurimoto:2006iv,Cheng:2014fwa,Lu:2000hj,Descotes-Genon:2001rya,Wei:2002iu} have demonstrated that $\phi_{B2}$ is not a negligible factor in HMEs, and its contributions to the form factor $F_0^{B \to \pi}$ in the PQCD approach can be as large as 30\% in certain cases \cite{Kurimoto:2006iv,Lu:2000hj}. Furthermore, the contribution of $\phi_{B2}$ to the branching ratio can be comparable to that from next-to-leading-order (NLO) corrections \cite{Yang:2020xal,Yang:2022ebu}. This paper focuses on the potential influence of $\phi_{B2}$ in $B_s \to PP$ and $PV$ decays within the PQCD framework. One commonly used expression for the leading $B_s$ mesonic wave function $\phi_B^+$ in actual PQCD calculations is given by \cite{Keum:2000wi},

    \begin{equation}
   {\phi}_{B}^{+}(x_{1},b_{1})\, =\, N\, x_{1}^{2}\,\bar{x}_{1}^{2}\,
   {\exp}\Big\{ -\Big( \frac{x_{1}\,m_{B_s}}{\sqrt{2}\,{\omega}_{B_s}}
    \Big)^{2} -\frac{1}{2} {\omega}_{B_s}^{2}\,b_{1}^{2} \Big\}
    \label{wf-b-meson-plus}.
    \end{equation}
   The corresponding $B_s$ mesonic WFs ${\phi}_{B}^{-}$
   is written as \cite{Yang:2020xal,Kurimoto:2006iv},
    \begin{align}
   {\phi}_{B}^{-}(x_{1},b_{1}) &= N\,
    \frac{2\,{\omega}_{B_s}^{4}}{m_{B_s}^{4}}\,
   {\exp}\Big(-\frac{1}{2} {\omega}_{B_s}^{2}\,b_{1}^{2} \Big)\,\Big\{
    \sqrt{{\pi}}\,\frac{m_{B_s}}{\sqrt{2}\,{\omega}_{B_s}}
   {\rm Erf}\Big( \frac{m_{B_s}}{\sqrt{2}\,{\omega}_{B_s}},
    \frac{x_{1}\,m_{B_s}}{\sqrt{2}\,{\omega}_{B_s}}\Big)
    \nonumber \\ & +
    \Big[1+\Big(\frac{m_{B_s}\,\bar{x}_{1}}{\sqrt{2}\,{\omega}_{B_s}}\Big)^{2}\Big]
   {\exp}\Big[-\Big(\frac{x_{1}\,m_{B_s}}{\sqrt{2}\,{\omega}_{B_s}}\Big)^{2} \Big]
   -{\exp}\Big(-\frac{m_{B_s}^{2}}{2\,{\omega}_{B_s}^{2}} \Big) \Big\}
    \label{wf-b-meson-minus}.
    \end{align}
   Here, ${\omega}_{B_s}$ denotes the shape parameter, and
   $\bar{x}_{1}$ $=$ $1$ $-$ $x_{1}$.
   The normalization constant $N_{B_s}$ is determined by the condition,
    \begin{equation}
   {\int}_{0}^{1}dx_{1}\, {\phi}_{B}^{\pm}(x_{1},0)\, =\, 1
    \label{wf-b-meson-normalization},
    \end{equation}
   which ensures that the wave function is properly normalized. The shapes of the $B_s$ meson DAs, including ${\phi}_{B}^{+}$ and ${\phi}_{B}^{-}$ are illustrated in Fig. \ref{fig-wf-1}.

 The following key observations can be made:
\begin{itemize}
  \item  The nonzero distributions of $\phi_B^\pm$ are predominantly located in the small $x$ region, and $\phi_B^\pm$ vanishes as $x \to 1$. This is consistent with the expectation that the light quark in the $B$-meson carries a small longitudinal momentum fraction.
   
  \item The shapes of $\phi_B^-$ and $\phi_B^+$ exhibit notable differences in the small $x$ region. Particularly, the DAs $\phi_B^-$ and $\phi_B^+$ show distinct endpoint behaviors at $x = 0$. This indicates that $\phi_{B2} = \phi_B^+ - \phi_B^- \neq 0$, suggesting that the approximation $\phi_{B2} = 0$ used in previous studies may be inappropriate and insufficient.

  \item Both $\phi_B^-$ and $\phi_{B2}$ are nonzero at the endpoint $x = 0$, resulting in an infrared divergence in the integral $\displaystyle \int \frac{dx}{x} \phi_{B2}$ under the collinear approximation. This suggests that the PQCD approach must account for nonperturbative contributions, potentially by considering the effects of the transverse momentum of the valence quarks and the inclusion of Sudakov factors.

  \item The distributions of $\phi_B^\pm$ are sensitive to the shape parameter $\omega_{B_s}$. As $\omega_{B_s}$ increases, the distributions of $\phi_B^\pm$ become broader. Consequently, the theoretical results with the PQCD framework depend on the chosen value of $\omega_{B_s}$.
\end{itemize}

   The WFs for the final states, including light pseudoscalar mesons and longitudinally polarized vector mesons, are defined as follows \cite{Ball:1998je,Ball:2006wn,Ball:2007rt,Kurimoto:2001zj},
   \begin{align} 
  &{\langle}\,P(p_{2})\,{\vert}\,\bar{q}_{i}(0)\,q_{j}(z)\,
  {\vert}\,0\,{\rangle}
   \nonumber \\ =&
  -i\,\frac{f_P}{4N_c}\,{\int}_{0}^{1}\mathbbm{d}x_{2}\,
   e^{+i\,k_{2}{\cdot}z}\,  \Big\{ {\gamma}_{5}\,
   \Big[ \!\!\not{p}_{2}\,{\phi}_{P}^{a}+
  {\mu}_{P}\,{\phi}_{P}^{p}-{\mu}_{P}\,
  (\not{n}_{-}\!\not{n}_{+}-1)\, {\phi}_{P}^{t}
   \Big] \Big\}_{ji},
   \\ 
  &{\langle}\,V(p_{3},e_{\parallel})\,{\vert}\,
   \bar{q}_{i}(0)\,q_{j}(z)\,{\vert}\,0\,{\rangle}
   \nonumber \\ =&
   \frac{1}{4N_c}\, {\int}_{0}^{1}\mathbbm{d}x_{3}\, e^{+i\,k_{3}{\cdot}z}\,
   \big\{ \!\not{e}_{{\parallel}}\,m_{V}\,f_{V}^{\parallel}\,
  {\phi}_{V}^{v}\,+ \!\not{e}_{{\parallel}}
   \!\not{p}_{3}\, f_{V}^{\perp}\, {\phi}_{V}^{t}- m_{V}\,
   f_{V}^{\perp}\, {\phi}_{V}^{s} \big\}_{ji}.
   \label{vector-meson-WF-definition},
   \end{align}
   In this context, $f_{P}$, $f_{V}^{\parallel}$, and $f_{V}^{\perp}$
  denote the decay constants. The wave functions ${\phi}_{P}^{a}$ and ${\phi}_{V}^{v}$ represent the twist-2 components, while
   ${\phi}_{P}^{p,t}$ and ${\phi}_{V}^{t,s}$ 
  correspond to the twist-3 components.
   
Following the conventions outlined in Refs.\cite{Ball:2006wn,Ball:2007rt}
and 
the DAs for the pseudoscalar meson $P$ $=$ $\pi,K$ meson are expressed as follows.
 \begin{align}
\phi_P^a(x)&=6x \bar{x}\{1+a_1^P C_1^{3/2}(\xi)+a_2^P C_2^{3/2}(\xi)\},\\
\phi_P^p(x)&=1+3\rho_+^P-9\rho_-^P a_1^P + 18\rho_+^P a_2^P \nonumber\\
             &+ \frac{3}{2}(\rho_+^P + \rho_-^P)(1- 3a_1^P +6a_2^P) \ln(x) \nonumber \\
             &+ \frac{3}{2}(\rho_+^P - \rho_-^P) (1+ 3a_1^P +6a_2^P) \ln(\bar{x}) \nonumber\\
             &- (\frac{3}{2} \rho_-^P - \frac{27}{2}\rho_+^P a_1^P + 27\rho_-^P a_2^P)C_1^{1/2}(\xi) \nonumber \\
             &+ (30\eta_p - 3\rho_-^P a_1^P +15\rho_+^P a_2^P)C_2^{1/2}(\xi), \\
\phi_P^t (x)&= \frac{3}{2}( \rho_-^P - 3\rho_+^P a_1^P + 6\rho_-^P a_2^P) \nonumber \\
             &-C_1^{1/2}(\xi) \{1 +3\rho_+^P - 12\rho_-^P a_1^P +24\rho_+^P a_2^P\nonumber \\
             &+\frac{3}{2}(\rho_+^P + \rho_-^P) (1- 3a_1^P +6a_2^P) \ln(x)  \nonumber\\
             &+\frac{3}{2}(\rho_+^P - \rho_-^P) (1+ 3a_1^P +6a_2^P) \ln(\bar{x})\nonumber   \\
             &-3(3\rho_+^P a_1^P - \frac{15}{2}\rho_-^P a_2^P )C_1^{1/2}(\xi).
\end{align}

Similarly, the DAs for the vector mesons are given by the following expressions

 \begin{align}
\phi_V^v(x)&= 6 x \bar{x}\left\{1+a_1^{\parallel,V} C_1^{3/2}(\xi)+a_2^{\parallel,V} C_2^{3/2}(\xi)\right\},\\
\phi_V^t (x)&=  3\xi\{C_1^{1/2}(\xi)+a_1^{\perp,V} C_2^{1/2}(\xi)+a_2^{\perp,V} C_3^{1/2}(\xi)\}\nonumber \\
 &+\frac{3}{2}\frac{m_s+m_q}{m_V}\frac{f_{V}^\parallel}{f_{V}^\perp} \{1+8\xi a_1^{\parallel,V}+(21-90x\bar{x})a_2^{\parallel,V}\nonumber \\
  &+\xi \ln{\bar{x}}(1+3a_1^{\parallel,V}+6a_2^{\parallel,V}) -\xi \ln{x}(1-3a_1^{\parallel,V}+6a_2^{\parallel,V})\}\nonumber  \\
             &-\frac{3}{2}\frac{m_s-m_q}{m_V}\frac{f_V^\parallel}{f_V^\perp}\xi \{2+9\xi a_1^{\parallel,V}+(22-60x\bar{x})a_2^{\parallel,V}\nonumber   \\
             &+ \ln{\bar{x}}(1+3a_1^{\parallel,V}+6a_2^{\parallel,V}) + \ln{x}(1-3a_1^{\parallel,V}+6a_2^{\parallel,V}) \},\\
\phi_V^s (x)&= \{-3C_1^{1/2}(\xi)-3 C_2^{1/2}(\xi)a_1^{\perp,V}-3 C_3^{1/2}(\xi)a_2^{\perp,V}\}\nonumber \\
             &-\frac{3}{2}\frac{m_s+m_q}{m_V}\frac{f_{V}^\parallel}{f_{V}^\perp} \{C_1^{1/2}(\xi)+2C_2^{1/2}(\xi)a_1^{\parallel,V}\nonumber \\
             &+[3C_3^{1/2}(\xi)+18C_1^{1/2}(\xi)]a_2^{\parallel,V}\nonumber  \\
             &+(\ln{\bar{x}}+1)(1+3a_1^{\parallel,V}+6a_2^{\parallel,V})\nonumber   \\
             &-(\ln{x}+1)(1-3a_1^{\parallel,V}+6a_2^{\parallel,V}) \}\nonumber   \\
             &+\frac{3}{2}\frac{m_s-m_q}{m_V}\frac{f_{V}^\parallel}{f_{V}^\perp} \{9C_1^{1/2}(\xi)a_1^{\parallel,V}+10C_2^{1/2}(\xi)a_2^{\parallel,V} \nonumber \\
             &+(\ln{\bar{x}}+1)(1+3a_1^{\parallel,V}+6a_2^{\parallel,V}) \}\nonumber   \\
             &+(\ln{x}+1)(1-3a_1^{\parallel,V}+6a_2^{\parallel,V}) \}.
\end{align}

In previous studies \cite{Yang:2020xal,Yang:2022ebu}, the effects of higher-order Gegenbauer polynomial terms - particularly those containing the third-order $ C_3^{m}(\xi) $ (which were only partially included) and the fourth-order $ C_4^{m}(\xi) $ polynomials - were not fully accounted for. To address this limitation, we systematically incorporate these higher-order contributions in our current framework. Following Refs.\cite{Ball:2007rt,Ball:2006wn}, we introduce modified wave functions (denoted with a prime) that extend the original formulations through the inclusion of additional terms derived from the complete set of higher-order Gegenbauer polynomials.

\begin{align}
\phi_P^{a'}(x)&=\phi_P^a(x)+6x \bar{x}\{a_4^P C_4^{3/2}(\xi)\},\\
\phi_P^{p'}(x)&=\phi_P^p(x) +(10\eta_p\lambda_{3p}-\frac{9}{2}\rho_-^pa_2^p)C_3^{1/2}(\xi) -3\eta_p\omega_{3p}C_4^{1/2}(\xi),\\
\phi_P^{t'} (x)&= \phi_P^t (x)
  -(5\eta_{P}-\frac{1}{2}\eta_{P}\omega_{3P}+\frac{3}{2}\rho_+^Ka_2^K)C_3^{1/2}-\frac{5}{3}\eta_P\lambda_{3P}C_4^{1/2},\\
  \phi_V^{v'}(x)&= \phi_{V}^v(x)+ 6 x \bar{x}\left\{a_4^{\parallel,V} C_4^{3/2}(\xi)\right\},\\
\phi_V^{t'} (x)&= \phi_V^{t} (x) +(15\kappa_{3V}^\perp-\frac{3}{2}\lambda_{3V}^\perp)C_3^{1/2}(\xi)+5\omega_{3V}^\perp C_4^{1/2}(\xi),\\
\phi_V^{s'} (x)&= \phi_V^{s} (x)+\{-15 C_2^{1/2}(\xi)\kappa_{3V}^\perp- C_3^{1/2}(\xi)\omega_{3V}^\perp\}+\frac{1}{4}\lambda_{3V}^\perp C_4^{1/2}(\xi).
\end{align}
Here, ${\xi}=x-\bar{x}=2\,x-1$, $\displaystyle \rho_+^P=\frac{M_P^2}{\mu_P^2}$, $\displaystyle \eta_P=\frac{f_{3P}}{f_P\mu_P}$. $C_{n}^{m}$ represents the Gegenbauer polynomials. $a_{n}^{P}$, $a_{n}^{{\parallel},V}$ and $a_{n}^{{\perp},V}$  are the corresponding the Gegenbauer moments.

\begin{figure}[H]
\centering
  ~~~~\includegraphics[width=4.5cm]{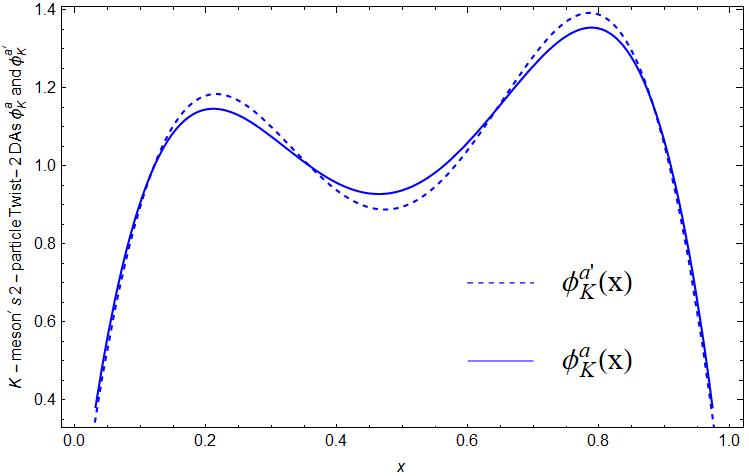}~~~\includegraphics[width=4.5cm]{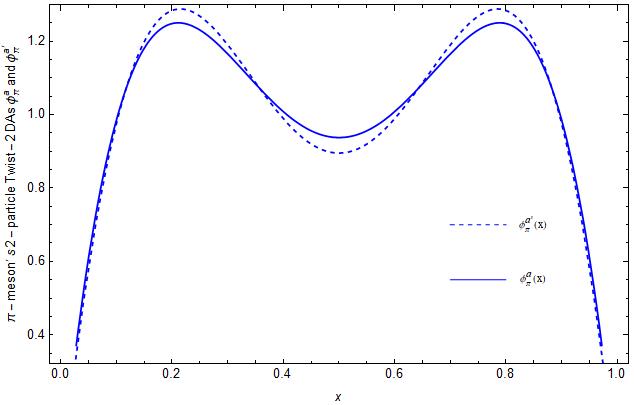}~~~\includegraphics[width=4.5cm]{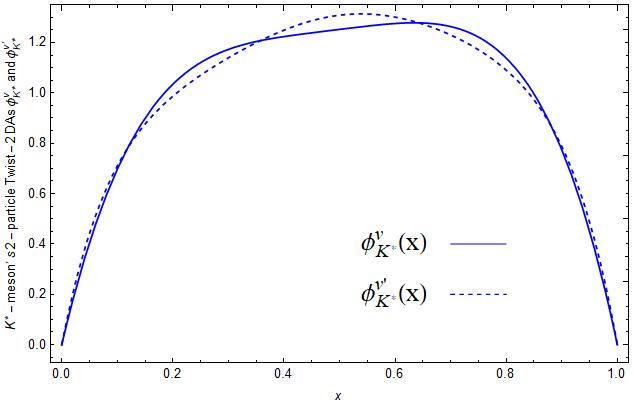}\\ \hspace{0.5cm}
  \includegraphics[width=4.5cm]{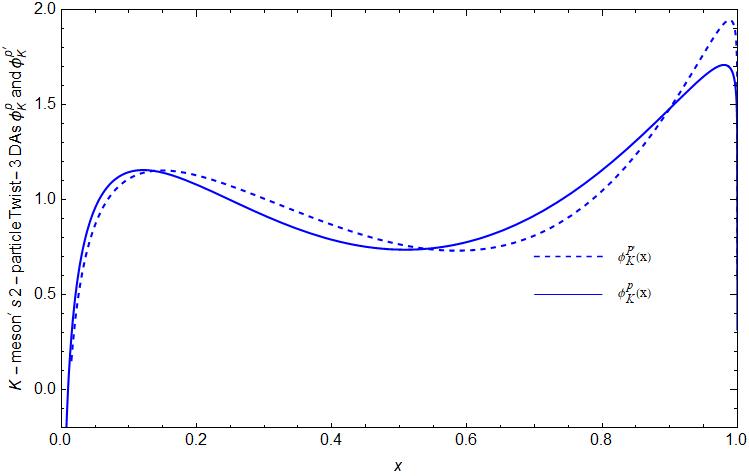}~~\includegraphics[width=4.5cm]{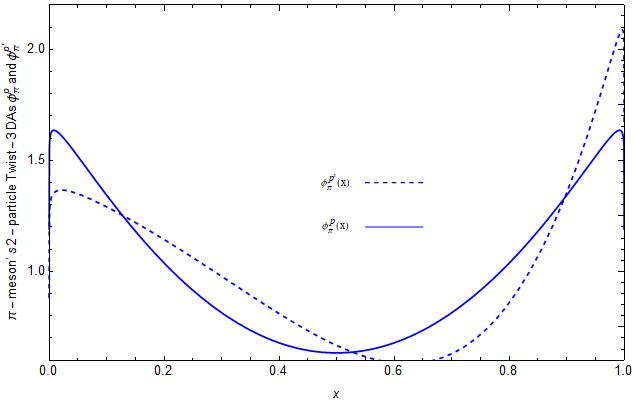}~~~\includegraphics[width=4.5cm]{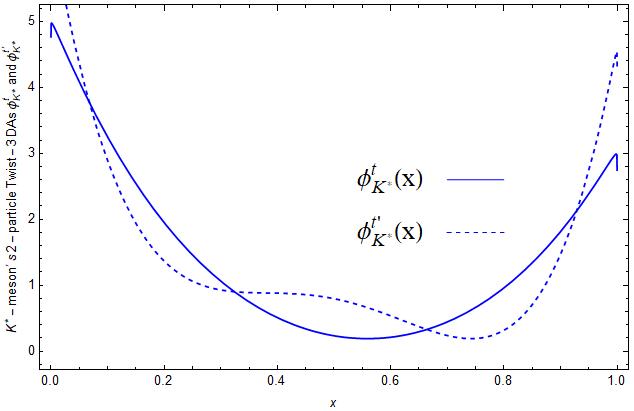}\\ \hspace{0cm}
  ~~~~\includegraphics[width=4.5cm]{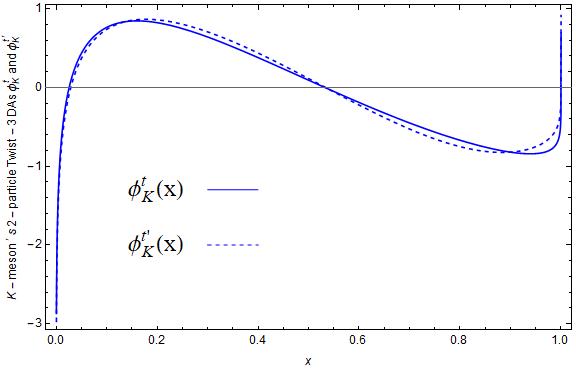}~\includegraphics[width=4.5cm]{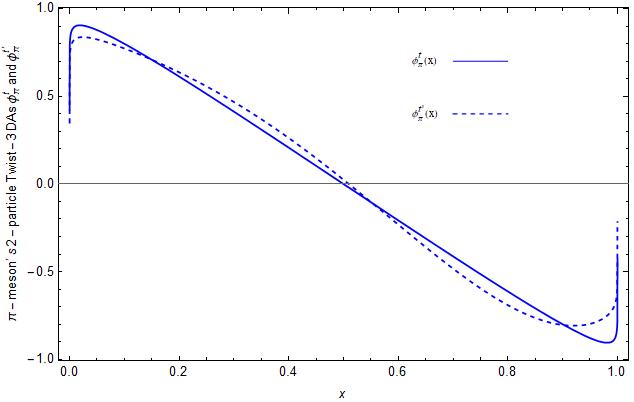}~~~\includegraphics[width=4.6cm]{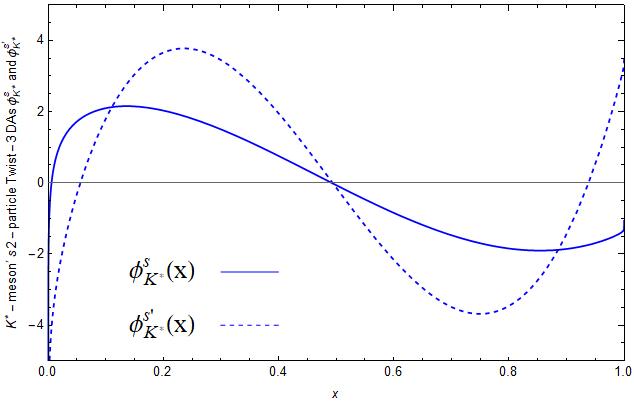}
  \caption{Shape lines of the final meson DAs, with (solid lines) and without (dotted lines) higher-order Gegenbauer polynomials, as a function of the longitudinal momentum fraction $x$.}
  \label{fig-wf-2}
\end{figure}

We present the DA curves for the $\pi$, $K$, and $K^*$ mesons as examples, shown in Fig.\ref{fig-wf-2}. From the figure, it is evident that higher-order terms influence the DAs; however, the exact extent of this effect requires further calculation.

\begin{table}[H]
\centering
   \caption{Summary of theoretical input parameters, with the central values taken as the default.}
   \begin{tabular}{cccc}
   \hline \hline
   \multicolumn{4}{c}{Wolfenstein parameters of the CKM matrix \cite{ParticleDataGroup:2024cfk}}\\ \hline
     $A=0.826^{+0.016}_{-0.015}$
   & ${\lambda}=0.22501\pm0.00068$
   & $\bar{\rho}=0.1591\pm0.0094$
   & $\bar{\eta}=0.3523^{+0.0073}_{-0.0071}$ \\ \hline
   \multicolumn{4}{c}{mass of particle (in the unit of MeV) \cite{ParticleDataGroup:2024cfk} (MeV)~ } \\ \hline
     $m_{{\pi}^{\pm}}=139.57$
   & $m_{K^{\pm}}=493.677\pm0.015$
   & $m_{\rho}=775.26\pm0.23$
   & $m_{K^{{\ast}{\pm}}}=895.5\pm0.8$  \\
     $m_{{\pi}^{0}}=134.98$
   & $m_{K^{0}}=497.611\pm0.013$
   & $m_{\omega}=782.66\pm0.13$
   & $m_{K^{{\ast}0}}=895.55\pm0.20$  \\
     $m_b=4780\pm6$
   & $m_{\phi}=1019.461\pm0.016$
   & $m_{B_{s}}=5366.93\pm0.10$
   & \\ \hline
   \multicolumn{4}{c}{decay constants (in the unit of MeV) and lifetime \cite{Ball:2007rt}
}
   \\ \hline
     $f_{\rho}^{\parallel}=216{\pm}3$ ~
   & $f_{\omega}^{\parallel}=187{\pm}5$~
   & $f_{\phi}^{\parallel}=215{\pm}5$ ~
   & $f_{K^{\ast}}^{\parallel}=220{\pm}5$~\\
     $f_{\rho}^{\perp}=165{\pm}9$ ~
   & $f_{\omega}^{\perp}=151{\pm}9$~
   & $f_{\phi}^{\perp}=186{\pm}9$ ~
   & $f_{K^{\ast}}^{\perp}=185{\pm}10$ ~  \\
     $f_{B_s}=230.3{\pm}1.3$\cite{ParticleDataGroup:2024cfk} ~
   & $f_{{\pi}}=130.2{\pm}0.8$\cite{ParticleDataGroup:2024cfk} ~
   & $f_{K}=155.7{\pm}0.3$\cite{ParticleDataGroup:2024cfk} ~ &$\tau_{B_{s}^0}=1.51$~ps~\\ \hline
   \multicolumn{4}{c}{Gegenbauer moments at the scale of
    ${\mu}=1$ GeV \cite{Ball:2006wn,Ball:2007rt}
 }
   \\ \hline
    $a_{1}^{{\pi}({\rho})}=a_{1}^{{\omega}({\phi})}=0$
   & $a_{2}^{\pi}=0.25{\pm}0.15$
   & $a_{1}^{K}=0.06{\pm}0.03$
   & $a_{2}^{K}=0.25{\pm}0.15$\\
     $a_{2}^{{\parallel},{\rho}({\omega})}=0.15{\pm}0.07$
   & $a_{2}^{{\parallel},{\phi}}=0.18{\pm}0.08$
   & $a_{1}^{{\parallel},K^{\ast}}= 0.03{\pm}0.02$
   & $a_{2}^{{\parallel},K^{\ast}} = 0.11{\pm}0.09$ \\
     $a_{2}^{{\perp},{\rho}({\omega})} = 0.14{\pm}0.06$
   & $a_{2}^{{\perp},{\phi}}=0.14{\pm}0.07$
   & $a_{1}^{{\perp},K^{\ast}}=0.04{\pm}0.03$
   & $a_{2}^{{\perp},K^{\ast}}=0.10{\pm}0.08$ \\

   $a_{4}^{{\rho}({\omega})}=0.03{\pm}0.02$
   & $a_{4}^{{\phi}}=0.02_{-0.02}^{+0.01}$
   & $a_{4}^{K^{\ast}}=0.02{\pm}0.01$
   & $a_{4}^{\pi,K}=-0.015$ \\ \hline \hline
   \label{tab:input}
   \end{tabular}
   \end{table}

\section{Numerical analysis \label{sec:numeric}}
\begin{figure}[tb]
\centering
  \includegraphics[width=14cm]{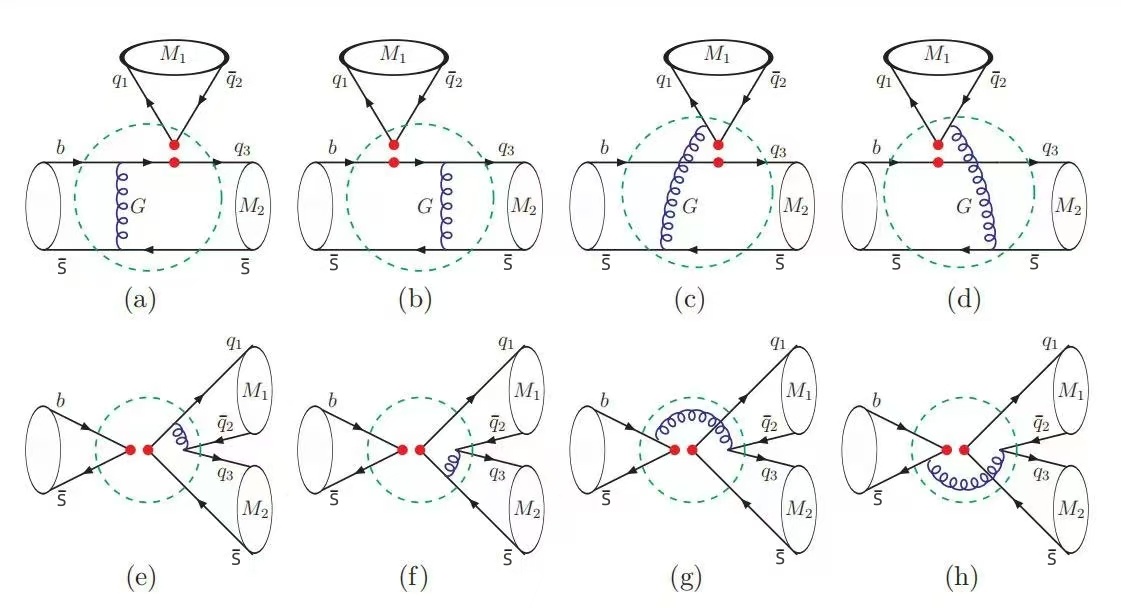}
  \caption{Feynman diagrams contributing to the $\bar{B}^0_s \to M_1 M_2$ decay processes. Dots denote interaction vertices, while the dashed circles represent quark-level scattering amplitudes. Diagrams are categorized as follows: (a) and (b) factorizable emission diagrams; (c) and (d) nonfactorizable emission diagrams; (e) and (f) factorizable annihilation diagrams; (g) and (h) nonfactorizable annihilation diagrams.  }
  \label{fig:abcd}
\end{figure}

The Feynman diagrams governing two-body nonleptonic $B_s$ meson decays are illustrated in Fig.\ref{fig:abcd}. Within the pQCD framework, the decay amplitude $\mathcal{A}$ is systematically factorized into three essential components: (i) short-distance Wilson coefficients $C_i$ encoding electroweak-scale dynamics, (ii) hard scattering kernels $\mathcal{H}_i$ describing quark-level interactions, and (iii) hadronic wave functions $\Phi_i$ parametrizing non-perturbative bound-state effects. The generalized amplitude structure can be expressed as 
  \begin{equation}
  \mathcal{A}_{i}\ {\propto}\ {\int}\, {\prod_j}dx_{j}\,db_{j}\,
  C_{i}(t_{i})\, \mathcal{H}_{i}(t_{i},x_{j},b_{j})\,
  {\Phi}_{j}(x_{j},b_{j})\,e^{-S_{j}}
  \label{hadronic-matrix-element}.
  \end{equation}
In the rest frame of the $B_s$ meson, the $CP$-averaged branching ratio is formally expressed as  
   \begin{equation}
   \mathcal{B}(B_s\to PP)\, =\,
   \frac{{\tau}_{B_s}}{16{\pi}}\,
   \frac{p_{\rm cm}}{m_{B_s}^{2}}\, \big\{
  {\vert}{\cal A}(B_s{\to}f){\vert}^{2}+
  {\vert}{\cal A}(\bar{B}_s{\to}\bar{f}){\vert}^{2} \big\}
   \label{ppbranching-ratio-definition},
   \end{equation}
    \begin{equation}
   \mathcal{B}(B_s\to PV)\, =\,
   \frac{{\tau}_{B_s}}{16{\pi}}\,
   \frac{p_{\rm cm}^3}{m_{V}^{2}}
   \frac{ 
  {\vert}{\cal A}(B_s{\to}f){\vert}^{2}+
  {\vert}{\cal A}(\bar{B}_s{\to}\bar{f}){\vert}^{2} }{(\epsilon \cdot p_{B_s})^2}
   \label{pvbranching-ratio-definition},
   \end{equation}   
where $ \tau_{B_s} $ denotes the $ B_s $-meson lifetime, $ p_{\rm cm} = \sqrt{\lambda(m_{B_s}^2, m_{M_1}^2, m_{M_2}^2)}/(2m_{B_s}) $ is the magnitude of the three-momentum shared by the pseudoscalar ($ P $) or vector ($ V $) mesons in the final state, and the  the Källén function $ \lambda(a,b,c) = a^2 + b^2 + c^2 - 2ab - 2ac - 2bc $. The explicit forms of the amplitudes, including $ \phi_{B2} $-dependent terms, are systematically derived in Appendix \ref{sec:mode}.

The time-dependent $CP$-violating asymmetry for the neutral $\bar{B}^0_{s}$ meson decay into a final state $f$ (with $f=\bar{f}$) is defined through the normalized decay rate difference:  
   \begin{equation}
   \mathcal{A}_{CP}\, =\,
   \frac{{\Gamma}(\bar{B}^{0}_s{\to}f)-{\Gamma}(B^{0}_s{\to}\bar{f})}
        {{\Gamma}(\bar{B}^{0}_s{\to}f)+{\Gamma}(B^{0}_s{\to}\bar{f})}
   \label{CP-asymmetry-definition-02}.
   \end{equation}
Notably, for the decay channel $\bar{B}_s^0 \to K^+ \pi^-$, the final state $\bar{f} = K^+ \pi^-$ is not a $CP$ eigenstate. Consequently, the direct $CP$ asymmetry is defined as
\begin{eqnarray}
{\cal A}_{\rm CP}(B_s^0 \to \pi^+K^-)= \frac{|\bar{A}_{\bar{f}}|^2 - |A_f|^2}{ |\bar{A}_{\bar{f}}|^2 + |A_f|^2 },
\label{eq:acp-1a}
\end{eqnarray}
where $A_f$ and $\bar{A}_{\bar{f}}$ denote the decay amplitudes for $B_s^0 \to f$ and $\bar{B}_s^0 \to \bar{f}$ processes, respectively.  


As indicated in the reference \cite{Yang:2020xal}, the selection of the parameter $\omega_{B_s}$ significantly influences the branching ratio calculations. To determine the optimal value of $\omega_{B_s}$, we employed the least-squares method, which is commonly used in theoretical physics to minimize the sum of the squares of the differences between observed and calculated values.

The $\chi^2$ statistic is defined as:

  \begin{equation}
  {\chi}^{2}~ =~ \sum\limits_{i}{\chi}^{2}_{i}~ =~ \sum\limits_{i}
   \frac{ ({\cal B}_{i}^{\rm th}-{\cal B}_{i}^{\rm exp})^{2} }
        { {\sigma}_{i}^{2} },
   \label{eq:chi2}
   \end{equation}
where ${\cal B}_i^{\rm th}$ and ${\cal B}_i^{\rm exp}$ represent the theoretical and experimental branching ratios for the $i$-th decay mode, respectively, and $\sigma_i$ denotes the experimental uncertainty.

\begin{figure}[H]
\centering
  \includegraphics[width=6cm]{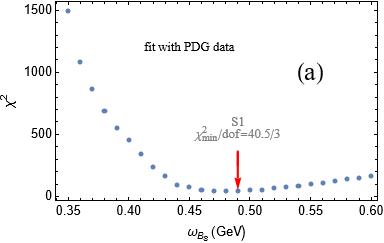}~~~\includegraphics[width=6cm]{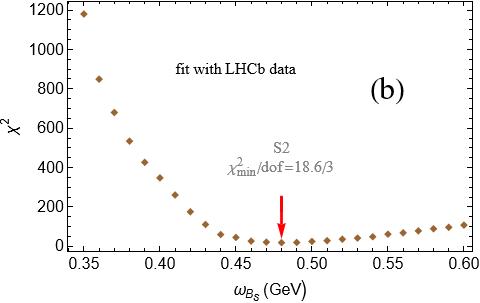}
  \caption{$\chi^2$ distribution as a function of the $B_s$-meson shape parameter $\omega_{B_s}$ for the $\bar{B}^0_s \to PP$ decay analysis. The global fit incorporates experimental constraints from $B_s \to PP$ measurements by PDG and LHCb. Red data points (marked by arrows) indicate the $\chi^2$ minima, corresponding to the optimized $\omega_{B_s}$ values.  }
  \label{fig:ppchi}
\end{figure}

\begin{figure}[H]
\centering
  \includegraphics[width=6cm]{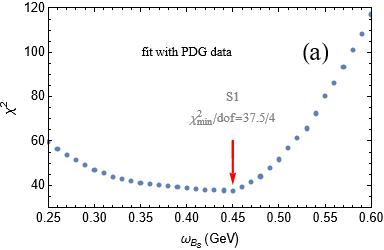}~~~\includegraphics[width=6cm]{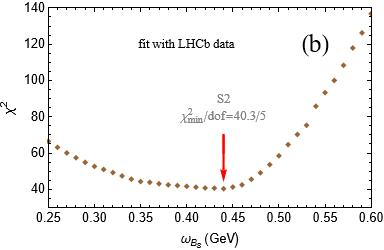}
  \caption{$\chi^2$ distribution as a function of the $B_s$-meson shape parameter $\omega_{B_s}$ for the $\bar{B}^0_s \to PV$ decay analysis. Other legends are the same as those of  figure \ref{fig:ppchi}.  }
  \label{fig:pvchi}
\end{figure}

\begin{figure}[H]
\centering
   \includegraphics[width=6cm]{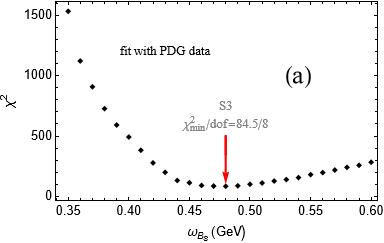}~~~\includegraphics[width=6cm]{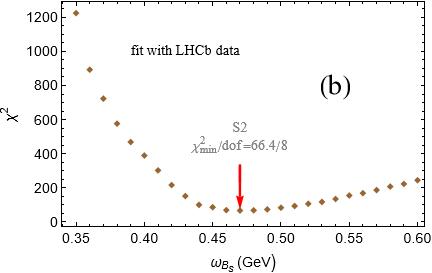}
  \caption{$\chi^2$ distribution as a function of the $B_s$-meson shape parameter $\omega_{B_s}$ for both  $\bar{B}^0_s \to PP$ and $PV$ decay analyses. Other legends are the same as those of  figure \ref{fig:ppchi}.   }
  \label{fig:allchi}
\end{figure}

We performed calculations for the decays $\bar{B}^0_s \to PP$ and $PV$ using data from both PDG and LHCb. The resulting $\chi^2$ distributions as functions of $\omega_{B_s}$ are presented in Figures \ref{fig:ppchi} and \ref{fig:pvchi} for the respective decay modes. The combined distribution, incorporating both decay processes, is shown in Figure \ref{fig:allchi}. The fitting results obtained from both datasets exhibit close agreement. However, given the more comprehensive nature of the PDG data, we selected the optimal values of $\omega_{B_s}$ derived from these calculations for further analysis:

\begin{itemize}
  \item For the decay $\bar{B}^0_s \to PP$, we adopted $\omega_{B_s} = 0.49$ GeV.
  \item For the decay $\bar{B}^0_s \to PV$, we selected $\omega_{B_s} = 0.45$ GeV.
  \item For the combined decay modes $\bar{B}^0_s \to PP, PV$, we chose $\omega_{B_s} = 0.48$ GeV.
\end{itemize}

The numerical results for the branching ratios of the decay modes $\bar{B}^0_s \to PP$ and $PV$ are shown in Tables \ref{b1br} and \ref{B1br}, respectively, while the direct $CP$ violation results are presented in Table \ref{b1cp} . Furthermore, a comparison between the branching ratios calculated with and without the inclusion of higher-order terms in the final-state meson DAs can be found in Table \ref{C3C4}. The results obtained for ${\omega}_{B_s} = 0.48$ GeV are listed in Table \ref{t11}, while the numerical results for the $\bar{B}^0_s \to PV$ decay mode from previous studies and the experimental data provided by PDG and HFLAV are shown in Table \ref{t05}.

From a comparison of the numerical results presented in the tables, we can draw the following conclusions:

\begin{itemize}
\item Branching Ratios: As indicated in Table \ref{b1br}, the branching ratio for the decay $B_s \to KK$ is generally larger than those for $B_s \to \pi K$ and $B_s \to \pi \pi$. For each decay channel, the inclusion of the contribution from $\phi_{B2}$ leads to a distinct enhancement in the branching ratio, particularly in certain decay channels (e.g., $\bar{B}_s^0 \to \pi^0 K^0$, $\bar{B}_s^0 \to K^{-} K^{+}$ and $\bar{B}_s^0 \to \bar{K}^0 K^0$), where the contribution from $\phi_{B2}$ can even reach the level of subleading corrections. This suggests that the impact of $\phi_{B2}$ cannot be neglected in precision calculations of $B_s$ meson weak decays.

\item Larger Branching Ratios: The larger branching ratios are observed for the decays $\bar{B}_s^0 \to K^{-} K^{+}$, $\bar{B}_s^0 \to \bar{K}^0 K^0$ and $\bar{B}_s^0 \to \rho^{-} K^{+}$. When both $\phi_{B1}$ and $\phi_{B2}$ contributions are considered, the branching ratios can reach magnitudes on the order of $10^{-5}$, which are likely to be observed experimentally in the future.

\item Direct CP Violation: Tables \ref{b1cp}  indicate that the contribution of $\phi_{B2}$ to direct CP violation varies across different decay channels. For instance, in the decay $\bar{B}_s^0 \to K^{-} K^{+}$, the contribution of $\phi_{B2}$ is $13\%$, while in the decays $\bar{B}_s^0 \to \pi^{+} \pi^{-}$ and $\bar{B}_s^0 \to \pi^{0} \pi^{0}$, the contribution of $\phi_{B2}$ is nearly a factor of two larger. Experimentally, the direct CP violation in the decay $\bar{B}_s^0 \to \pi^{-} K^{+}$ is measured to be $0.224 \pm 0.012$, and our calculated value is of the same order of magnitude. Since CP violation is closely related to strong phase shifts, obtaining precise information about these strong phases is essential for a more comprehensive understanding of CP violation phenomena. However, the factors influencing the strong phases, such as higher-order radiative corrections to hadronic matrix elements and final-state interactions between particles, complicate the analysis. Additionally, due to the lack of effective constraints on these factors, the experimental precision of CP violation measurements is often limited, resulting in a sizable discrepancy between theoretical predictions and experimental data.

\item Uncertainty Analysis: The impact of various theoretical parameters on the branching ratio is discussed in Tables \ref{b1br} and \ref{B1br}. Among the parameters, the Gegenbauer parameter, chiral masses, and the wave function parameter $\omega_{B_s}$ exhibit significant effects, while the uncertainty due to the $B_s$ meson decay constant is relatively small. The error due to variations in $\mu_P$ ranges from $5.0\%$ to $32.0\%$, while the error due to changes in $\omega_{B_s}$ ranges from $2.8\%$ to $20.0\%$. The error due to the Gegenbauer parameter is stable, typically between $7\%$ and $10\%$, and the total theoretical uncertainty falls within the range of $9.2\%$ to $33.5\%$. Given the sensitivity of the branching ratio to the wave function parameter $\omega_{B_s}$, an appropriate choice of $\omega_{B_s}$ can enhance the contribution from $\phi_{B1}$ and partially compensate for the contribution from $\phi_{B2}$. This may explain why the contribution from $\phi_{B2}$ has often been overlooked in previous studies. Overall, $\phi_{B2}$ does affect both the branching ratios and CP violation, and its contribution should not be ignored.

\item Higher-Order Terms: As seen in Table \ref{C3C4}, the inclusion of higher-order terms in the final-state meson DAs leads to changes in the branching ratio calculations. For example, the inclusion of higher-order terms in the decay $\bar{B}_s^0 \to \rho^0 K^0$ results in an impact on the branching ratio of up to $6\%$. Overall, within the allowed error range, our calculations are consistent with previous theoretical predictions, although the inclusion of higher-order terms and the different choices of parameters contribute to some deviations in the results.

\item Fitting with Multiple Decay Channels: When both decay modes are considered simultaneously in the fit, the optimal value of $\omega_{B_s}$ is found to be 0.48 GeV, which is consistent with results from references \cite{Hua:2020usv,Keum:2000wi}. Using this optimal $\omega_{B_s}$, we obtain the results presented in Table \ref{t11}. Compared to previous calculations, we observe that the branching ratios for some decay modes increase with an increasing $\omega_{B_s}$, while for others, they decrease. From a comprehensive analysis of all the tables, we infer that previous calculations likely selected an appropriate $\omega_{B_s}$ to minimize the discrepancy between theoretical predictions and experimental data. This suggests that the contribution from $\phi_{B2}$ may be effectively replaced by other input parameters, which could explain why the contribution from $\phi_{B2}$ has often been neglected in earlier studies.

\end{itemize}


 \begin{table}[H]
  \centering
  \caption{\small CP-averaged branching ratios (in units of $10^{-6}$) for $\bar{B}^0_s \to PP$ decays. For each decay channel, the first row corresponds to the results with $\phi_{B_2}=0$, while the second row shows the case with $\phi_{B_2}\neq0$. The systematic uncertainties in columns 3-6 are respectively induced by the Gegenbauer parameter $a_2^P = 0.25 \pm 0.15$, factorization scale $\mu_{P} = 1.4 \pm 0.1~\mathrm{GeV}$, and $B_s$-meson shape parameter $\omega_{B_s} = 0.49 \pm 0.01~\mathrm{GeV}$ and decay constant $f_{B_s}=230.3\pm1.3$~MeV (see Table \ref{tab:input} for other input parameters). }
  \begin{tabular}{ccccccc}
  \hline\hline
    mode &Center value
    & $\rm{a_2^{P}}$
         & $\rm{\mu_P}$
         & $\rm{\omega_{B_s}}$
         &$f_{B_s}$
          & $\rm{total}$\\ \hline
   \multirow{2}{*}{$\bar{B}_s^0 \rightarrow \pi^{-} K^{+}$} &$6.599$
  & $ _{-0.495}^{+0.572}$ & $^{+0.756}_{-0.562}$ & $_{-0.436}^{+0.485}$
    & $^{+0.128}_{-0.127}$& $^{+1.073}_{-0.876}$\\ 
    &$8.639$
  & $ _{-0.397}^{+0.453}$ & $^{+0.880}_{-0.692}$ & $_{-0.307}^{+0.242}$
   &  $^{+0.104}_{-0.121}$& $^{+1.102}_{-0.863}$\\ \hline
    \multirow{2}{*}{$ \bar{B}_s^0 \rightarrow \pi^{-} \pi^{+}$}&$0.634$
  & $_{-0.050}^{+0.055} $ & $_{-0.109}^{+0.166}$ & $_{-0.029}^{+0.102}$
& $^{+0.033}_{-0.032}$& $^{+0.205}_{-0.127}$\\
&$0.958$ & $_{-0.085}^{+0.096} $ & $_{-0.127}^{+0.134}$ & $_{-0.063}^{+0.091}$
&  $^{+0.054}_{-0.051}$& $^{+0.196}_{-0.173}$\\ \hline
    \multirow{2}{*}{$ \bar{B}_s^0 \rightarrow \pi^{0} K^{0}$}&$0.170$
  & $_{-0.014}^{+0.012}$ & $^{+0.021}_{-0.024}$ & $_{-0.009}^{+0.008}$
 &  $^{+0.007}_{-0.007}$& $^{+0.026}_{-0.030}$\\
 &$0.260$ & $_{-0.021}^{+0.022}$ & $^{+0.040}_{-0.044}$ & $_{-0.052}^{+0.046}$
 & $^{+0.010}_{-0.010}$& $^{+0.066}_{-0.072}$\\ \hline
  \multirow{2}{*}{ $ \bar{B}_s^0 \rightarrow K^{-} K^{+}$}&$12.997$
  & $_{-1.212}^{+1.134}$ & $^{+3.088}_{-3.559}$ & $_{-1.506}^{+0.130}$
&  $^{+0.062}_{-0.089}$& $^{+3.293}_{-4.051}$\\
&$19.040$& $_{-1.458}^{+1.327}$ & $^{+3.930}_{-4.423}$ & $_{-2.708}^{+1.604}$
&  $^{+0.429}_{-0.424}$& $^{+4.468}_{-5.404}$\\ \hline
   \multirow{2}{*}{ $ \bar{B}_s^0 \rightarrow K^0 \bar{K}^{0}$}&$15.190$
  & $_{-1.411}^{+1.394} $ & $^{+2.032}_{-1.970}$ & $_{-1.648}^{+0.93}$ &  $^{+0.349}_{-0.342}$& $^{+2.657}_{-2.950}$\\
      &$18.502$  & $_{-1.582}^{+1.653} $ & $^{+1.984}_{-1.887}$ & $_{-2.092}^{+1.064}$  & $^{+0.427}_{-0.416}$& $^{+2.825}_{-3.258}$\\ \hline
   \multirow{2}{*}{$ \bar{B}_s^0 \rightarrow \pi^{0} \pi^{0}$}&$0.317$
  & $_{-0.025}^{+0.028}$ & $^{+0.083}_{-0.055}$ & $_{-0.014}^{+0.051}$ & $^{+0.017}_{-0.016}$& $^{+0.103}_{-0.064}$\\
   &$0.479$& $_{-0.039}^{+0.045}$ & $^{+0.067}_{-0.064}$ & $_{-0.031}^{+0.048}$  & $^{+0.027}_{-0.026}$& $^{+0.098}_{-0.085}$\\
  \hline\hline
  \end{tabular}
  \label{b1br}
  \end{table}
   \begin{table}[H]
  \centering
  \caption{\small   CP-averaged branching ratios (in units of $10^{-6}$) for $\bar{B}^0_s \to PV$ decays, $B_s$-meson shape parameter $\omega_{B_s} = 0.45 \pm 0.01~\mathrm{GeV}$. Other notations and theoretical error conventions follow  Table \ref{b1br} . }
  \begin{tabular}{cccccccc}
  \hline\hline
    mode &Center value& $\rm{a_2^{P}}$
         & $\rm{\mu_P}$
         & $\rm{\omega_{B_s}}$
         &$f_{B_s}$
          & $\rm{total}$\\ \hline
    \multirow{2}{*}{ $ \bar{B}_s^0 \rightarrow \rho^0K^{0} $}&$0.088$
  & $ ^{+0.007}_{-0.007}$ & $^{+0.007}_{-0.006}$ & $_{-0.008}^{+0.013}$
   &    $^{+0.004}_{-0.004}$ &  $^{+0.017}_{-0.013}$  \\
    &$0.089$
  & $ ^{+0.007}_{-0.008}$ & $^{+0.009}_{-0.008}$ & $_{-0.007}^{+0.013}$
    &  $^{+0.004}_{-0.004}$ &  $^{+0.018}_{-0.014}$\\ \hline
     \multirow{2}{*}{ $\bar{B}_s^0 \rightarrow \pi^{0} K^{*0}$}&$0.108$
  & $_{-0.009}^{+0.009} $ & $_{-0.008}^{+0.007}$ & $_{-0.003}^{+0.002}$
  &     $^{+0.001}_{-0.002}$ &  $^{+0.012}_{-0.013}$ \\
  &$0.126$
  & $_{-0.012}^{+0.011} $ & $_{-0.009}^{+0.009}$ & $_{-0.003}^{+0.003}$
    &  $^{+0.001}_{-0.002}$&  $^{+0.015}_{-0.015}$ \\ \hline
    \multirow{2}{*}{  $ \bar{B}_s^0 \rightarrow \omega K^{0} $}&$0.106$
  & $_{-0.008}^{+0.008}$ & $^{+0.025}_{-0.025}$ & $_{-0.013}^{+0.013}$
  &    $^{+0.009}_{-0.008}$&  $^{+0.031}_{-0.030}$ \\
  &$0.169$
  & $_{-0.013}^{+0.014}$ & $^{+0.027}_{-0.029}$ & $_{-0.019}^{+0.017}$
   &  $^{+0.010}_{-0.009}$&  $^{+0.036}_{-0.038}$\\ \hline
    \multirow{2}{*}{ $  \bar{B}_s^0 \rightarrow K^{0} \phi$}&$0.153$
  & $_{-0.011}^{+0.013}$ & $^{+0.011}_{-0.016}$ & $_{-0.001}^{+0.001}$
&     $^{+0.005}_{-0.005}$&  $^{+0.018}_{-0.020}$  \\
&$0.147$
  & $_{-0.011}^{+0.012}$ & $^{+0.010}_{-0.015}$ & $_{-0.001}^{+0.001}$
   &  $^{+0.006}_{-0.006}$ &  $^{+0.017}_{-0.020}$\\ \hline
     \multirow{2}{*}{ $ \bar{B}_s^0 \rightarrow K^{*-} K^{+}$}&$8.196$
  & $_{-0.694}^{+0.725} $ & $^{+0.433}_{-0.427}$ & $_{-0.235}^{+0.224}$ &   $^{+0.104}_{-0.091}$ &  $^{+0.880}_{-0.853}$ \\
   &$11.996$
  & $_{-1.096}^{+1.125} $ & $^{+0.551}_{-0.425}$ & $_{-0.337}^{+0.331}$  &  $^{+0.137}_{-0.164}$ &  $^{+1.303}_{-1.234}$ 
  \\ \hline
    \multirow{2}{*}{ $ \bar{B}_s^0 \rightarrow  K^{-}K^{*+}$}&$4.889$
  & $_{-0.280}^{+0.339}$ & $^{+0.333}_{-0.314}$ & $_{-0.156}^{+0.199}$ &  $^{+0.014}_{-0.019}$&  $^{+0.515}_{-0.449}$ \\
  &$7.584$
  & $_{-0.734}^{+0.692}$ & $^{+0.538}_{-0.561}$ & $_{-0.307}^{+0.352}$  &  $^{+0.021}_{-0.023}$&  $^{+0.945}_{-0.974}$
  \\ \hline
    \multirow{2}{*}{ $ \bar{B}_s^0 \rightarrow \bar{K}^{*0}K^{0} $}&$6.808$
  & $ _{-0.453}^{+0.492}$ & $^{+0.550}_{-0.562}$ & $_{-0.546}^{+1.088}$
    &  $^{+0.021}_{-0.021}$&  $^{+1.315}_{-0.905}$ \\
   &$9.832$
  & $ _{-0.892}^{+0.877}$ & $^{+0.866}_{-0.887}$ & $_{-1.415}^{+1.709}$
    &  $^{+0.025}_{-0.025}$&  $^{+2.107}_{-1.893}$\\ \hline
    \multirow{2}{*}{  $ \bar{B}_s^0 \rightarrow \bar{K}^{0} K^{*0}$}&$5.480$
  & $  _{-0.398}^{+0.433}$ & $^{+0.830}_{-1.135}$ & $_{-0.352}^{+0.330}$
    &  $^{+0.017}_{-0.020}$&  $^{+0.993}_{-1.253}$ \\
   &$6.626$
  & $  _{-0.593}^{+0.635}$ & $^{+1.648}_{-2.123}$ & $_{-0.293}^{+0.271}$
    &  $^{+0.023}_{-0.023}$&  $^{+1.787}_{-2.224}$\\  \hline
     \multirow{2}{*}{ $ \bar{B}_s^0 \rightarrow \pi^{-} K^{*+}$}&$6.890$
  & $ _{-0.623}^{+0.580}$ & $^{+0.376}_{-0.275}$ & $ _{-0.438}^{+0.407}$
     &  $^{+0.015}_{-0.014}$&  $^{+0.802}_{-0.810}$  \\
      &$7.782$
  & $ _{-0.690}^{+0.745}$ & $^{+0.125}_{-0.129}$ & $ _{-0.486}^{+0.406}$
     &  $^{+0.010}_{-0.015}$&  $^{+0.858}_{-0.854}$ \\  \hline
     \multirow{2}{*}{ $  \bar{B}_s^0 \rightarrow \rho^{-}K^{+} $}&$16.966$
  & $  _{-1.253}^{+1.156} $ & $^{+1.014}_{-1.015}$ & $_{-1.103}^{+1.117}$
    &  $^{+0.229}_{-0.171}$&  $^{+1.914}_{-1.961}$ \\
   &$17.207$
  & $  _{-1.420}^{+1.359} $ & $^{+1.024}_{-1.025}$ & $_{-1.118}^{+1.414}$
    &  $^{+0.246}_{-0.209}$&  $^{+2.226}_{-2.088}$\\
  \hline\hline
  \end{tabular}
  \label{B1br}
  \end{table}
\begin{table}[H]
 \caption{\small The CP asymmetries $\mathcal{A}_{CP}$ (in the unit of $10^{-2}$) for the $\bar{B}^0_s \to PP$ and $\bar{B}^0_s \to PV$   decays. Notations and theoretical error conventions follow Tables  \ref{b1br} and \ref{B1br}.}
    \begin{center}
   \begin{tabular}{ccccccccc}
      \hline\hline
        mode&Center value & $\rm{a_2^{P}}$
         & $\rm{\mu_P}$
         & $\rm{\omega_{B_s}}$
         &$f_{B_s}$
          & $\rm{total}$\\ \hline
     \multirow{2}{*}{$ \bar{B}_s^0 \rightarrow \pi^{-} K^{+}$}&18.325
  & $ _{-2.341}^{+2.337}$ & $^{+0.626}_{-0.429}$ & $_{-1.018}^{+0.069}$
    & $^{+0.056}_{-0.057}$& $^{+2.421}_{-2.589}$\\
   &16.232
  & $ _{-2.163}^{+2.159}$ & $^{+0.504}_{-2.433}$ & $_{-0.626}^{+0.421}$
    & $^{+0.011}_{-0.009}$& $^{+2.257}_{-3.315}$\\ \hline
      \multirow{2}{*}{$ \bar{B}_s^0 \rightarrow \pi^{-} \pi^{+}$}&$-0.917$
  & $_{-0.376}^{+0.374} $ & $_{-0.057}^{+0.056}$ & $_{-0.014}^{+0.199}$
 & $-$& $^{+0.427}_{-0.381}$\\
    &$-1.776$
  & $_{-0.383}^{+0.382} $ & $_{-0.016}^{+0.015}$ & $_{-0.009}^{+0.167}$
 & $-$& $^{+0.417}_{-0.383}$\\ \hline
     \multirow{2}{*}{ $ \bar{B}_s^0 \rightarrow \pi^{0} K^{0}$}&32.089
  & $_{-4.263}^{+4.492}$ & $^{+0.575}_{-0.364}$ & $_{-0.942}^{+0.992}$
  & $-$& $^{+4.636}_{-4.381}$\\
    &42.858
  & $_{-4.749}^{+4.936}$ & $^{+1.066}_{-0.870}$ &$_{-0.889}^{+1.853}$
  & $-$& $^{+5.764}_{-5.373}$\\ \hline
     \multirow{2}{*}{$ \bar{B}_s^0 \rightarrow K^{-} K^{+}$}&$-4.808$
  & $_{-1.597}^{+1.236}$ & $^{+0.425}_{-0.230}$ & $_{-0.566}^{+0.717}$
 & $^{+0.003}_{-0.002}$& $^{+1.509}_{-1.725}$\\
   &$-4.179$
  & $_{-1.458}^{+1.327}$ & $^{+0.875}_{-0.428}$ & $_{-0.449}^{+0.535}$
 & $^{+0.002}_{-0.001}$& $^{+2.016}_{-2.089}$\\ \hline
    \multirow{2}{*}{ $ \bar{B}_s^0 \rightarrow \pi^{0} \pi^{0}$}&$-0.917$
  & $_{-0.376}^{+0.374}$ & $^{+0.056}_{-0.067}$ & $_{-0.014}^{+0.199}$  & $-$& $^{+0.429}_{-0.382}$\\
    &$-1.776$
  & $_{-0.383}^{+0.382}$ & $^{+0.015}_{-0.016}$ & $_{-0.009}^{+0.167}$  & $-$& $^{+0.421}_{-0.388}$\\ \hline
   \multirow{2}{*}{ $  \bar{B}_s^0 \rightarrow \rho^0K^{0} $}&71.748
  & $ ^{+9.334}_{-5.298}$ & $^{+7.859}_{-2.326}$ & $_{-0.061}^{+0.200}$
     &  $^{+0.046}_{-0.012}$ &  $^{+12.204}_{-5.786}$  \\
      &70.823
  & $ ^{+8.276}_{-5.131}$ & $^{+8.468}_{-2.050}$ & $_{-0.068}^{+0.205}$
    &  $^{+0.020}_{-0.013}$ &  $^{+11.842}_{-5.526}$  \\ \hline
     \multirow{2}{*}{  $\bar{B}_s^0 \rightarrow \pi^{0} K^{*0}$}&$-26.540$
  & $_{-5.363}^{+5.359} $ & $_{-1.599}^{+4.248}$ & $_{-2.517}^{+0.549}$
     &  $^{+0.187}_{-0.283}$ &  $^{+6.863}_{-6.143}$ \\
        &$-37.73$5
  & $_{-6.763}^{+6.286} $ & $_{-2.315}^{+5.655}$ & $_{-4.052}^{+1.564}$
     &  $^{+0.166}_{-0.315}$ &  $^{+8.600}_{-8.223}$ \\ \hline
     \multirow{2}{*}{  $ \bar{B}_s^0 \rightarrow \omega K^{0} $}&33.740
  & $_{-4.039}^{+3.798}$ & $^{+1.435}_{-1.786}$ & $_{-5.599}^{+4.563}$
    &  $^{+0.165}_{-0.155}$&  $^{+6.110}_{-7.133}$ \\
    &46.638
  & $_{-5.557}^{+5.296}$ & $^{+0.078}_{-0.066}$ & $_{-5.415}^{+3.877}$
   &  $^{+0.177}_{-0.169}$&  $^{+6.566}_{-7.761}$ \\ \hline
  \multirow{2}{*}{ $ \bar{B}_s^0 \rightarrow K^{*-} K^{+}$}&$-17.482$
  & $_{-2.649}^{+2.235} $ & $^{+0.340}_{-0.429}$ & $^{+0.384}_{-0.238}$
   &  $^{+0.098}_{-0.042}$ &  $^{+2.295}_{-2.694}$ \\
    &$-12.551$
  & $_{-2.253}^{+2.012} $ & $^{+0.325}_{-0.446}$ & $^{+1.083}_{-0.446}$
    &  $^{+0.031}_{-0.020}$ &  $^{+2.308}_{-2.340}$ \\ \hline
     \multirow{2}{*}{ $ \bar{B}_s^0 \rightarrow K^{-}K^{*+} $}&10.586
  & $_{-7.379}^{+7.253}$ & $^{+5.181}_{-6.836}$ & $_{-0.429}^{+0.340}$
   &  $^{+0.069}_{-0.026}$&  $^{+8.920}_{-10.068}$ \\
  &8.948
  & $_{-6.346}^{+6.203}$ & $^{+0.282}_{-0.986}$ & $_{-0.407}^{+0.326}$
   &  $^{+0.046}_{-0.073}$&  $^{+6.218}_{-6.435}$ \\ \hline
   \multirow{2}{*}{  $ \bar{B}_s^0 \rightarrow \pi^{-} K^{*+}$}&5.774
  & $ _{-2.752}^{+2.737}$ & $^{+2.145}_{-1.414}$ & $ _{-0.696}^{+0.693}$
     &  $^{+0.175}_{-0.183}$&  $^{+3.550}_{-3.177}$  \\
    &$-2.051$
  & $ _{-1.846}^{+1.839}$ & $^{+1.718}_{-0.756}$ & $ _{-0.915}^{+0.918}$
    &  $^{+0.195}_{-0.207}$&  $^{+2.686}_{-2.204}$  \\ \hline
     \multirow{2}{*}{  $  \bar{B}_s^0 \rightarrow \rho^{-}K^{+} $}&$-34.159$
  & $  _{-4.393}^{+4.460} $ & $^{+2.857}_{-2.958}$ & $_{-0.068}^{+0.085}$
    &  $^{+0.022}_{-0.015}$&  $^{+5.297}_{-5.297}$ \\
     &$-35.164$
  & $  _{-5.326}^{+5.285} $ & $^{+3.521}_{-3.506}$ & $_{-0.048}^{+0.024}$
    &  $^{+0.009}_{-0.003}$&  $^{+6.351}_{-6.377}$ \\
\hline\hline
\label{b1cp}
  \end{tabular}
   \end{center}
  \end{table}
\begin{table}[H]
\centering
   \caption{\small Comparison of branching ratio (in the unit of $10^{-6}$) results after incorporating higher-order terms into the final-state meson DAs. Symbols $\phi_{B1}'$ and $\phi_{B2}'$ denote predictions including these corrections, while $\omega_{B_s}$ is set to $0.49$ GeV for $PP$ and $0.45$ GeV for $PV$.}
   \begin{tabular}{cc|cc|cc|cc|cc}
      \hline\hline
        & Mode
        & \multicolumn{2}{c|}{$ \bar{B}_s^0 \rightarrow \pi^{-} K^{+}$}
          & \multicolumn{2}{c|}{$ \bar{B}_s^0 \rightarrow \pi^{+}\pi^{-}$}
         & \multicolumn{2}{c|}{ $ \bar{B}_s^0 \rightarrow \pi^{0} K^{0}$}
         & \multicolumn{2}{c}{ $ \bar{B}_s^0 \rightarrow K^{-} K^{+}$}
        \\  \hline
    & $\phi_{B1}/\phi_{B1'}$
         &  $6.599$&$6.738$
        & $0.634$&$0.652$
        &$ 0.170$&$0.175$
        &$ 12.997$&$12.744$\\
  & $\phi_{B2}/\phi_{B2'}$
        & 8.639&8.971
        & 0.958&1.009
        & 0.260&0.273
        & 19.040&18.589

   \\  \hline\hline

            & Mode
            & \multicolumn{2}{c|}{$ \bar{B}_s^0 \rightarrow \bar{K}^{0}K^0$}
          & \multicolumn{2}{c|}{$ \bar{B}_s^0 \rightarrow \pi^{0} \pi^{0}$}
         & \multicolumn{2}{c|}{$ \bar{B}_s^0 \rightarrow \rho^0 K^{0} $}
         & \multicolumn{2}{c}{ $ \bar{B}_s^0 \rightarrow \pi^{0} K^{*0}$}
        \\  \hline
    & $\phi_{B1}/\phi_{B1'}$
        & 15.190&15.305
        & 0.317&0.326
        & 0.088&0.082
        & 0.108&0.109\\
  & $\phi_{B2}/\phi_{B2'}$
        & 18.502&18.669
        & 0.479&0.505
        & 0.089&0.083
        & 0.126&0.120
          \\ \hline\hline       & Mode
        &\multicolumn{2}{c|}{$ \bar{B}_s^0 \rightarrow \omega K^{0}$}
        &\multicolumn{2}{c|}{$ \bar{B}_s^0 \rightarrow K^{0} \phi$}
        & \multicolumn{2}{c|}{$ \bar{B}_s^0 \rightarrow K^{*-} K^{+}$}
        &\multicolumn{2}{c}{$ \bar{B}_s^0 \rightarrow K^{-}K^{*+}$}
        \\  \hline
    & $\phi_{B1}/\phi_{B1'}$
        & $0.106$&$0.109$
        & $0.153$&$0.160$
        & $8.196$& $8.562$
        & $4.889$&$5.134$\\
   & $\phi_{B2}/\phi_{B2'}$
        & 0.169&0.174
        & 0.147&0.149
        &11.996&12.531
        & 7.584&7.815
          \\ \hline\hline & Mode
        & \multicolumn{2}{c|}{$ \bar{B}_s^0 \rightarrow \bar{K}^{*0}K^{0}$}
        &\multicolumn{2}{c|}{$ \bar{B}_s^0 \rightarrow \bar{K}^{0} K^{*0}$}
        &\multicolumn{2}{c|}{$ \bar{B}_s^0 \rightarrow \pi^{-} K^{*+}$}
        &\multicolumn{2}{c}{$ \bar{B}_s^0 \rightarrow \rho^{-}K^{+}$}
        \\  \hline
    & $\phi_{B1}/\phi_{B1'}$
         &  $6.808$& $7.049$
        & $5.480$&$5.673$
        &$ 6.890$&$7.128$
        &$16.966$ &$17.175$\\
  & $\phi_{B2}/\phi_{B2'}$
        &9.832&10.116
        & 6.626&7.039
        & 7.782&8.226
        & 17.207&18.026\\
          \hline\hline
  \end{tabular}
  \label{C3C4}
  \end{table}
\begin{table}[H]
\centering
    \caption{\small 
    Branching ratios (in units of $10 ^{-6}$)  for the  $\bar{B}_s^0 \to PP,PV$ decays with $\omega_{B_s}=0.48\rm{GeV}$.}
   \begin{tabular}{cccccccc}
      \hline\hline
        & Mode
        &$ \bar{B}_s^0 \rightarrow \pi^{-} K^{+}$
        & $ \bar{B}_s^0 \rightarrow \pi^{+}\pi^{-}$
        & $ \bar{B}_s^0 \rightarrow \pi^{0} K^{0}$
        & $ \bar{B}_s^0 \rightarrow K^{-} K^{+}$
        \\  \hline
    & $\phi_{B1}+{\phi}_{B2}$
         &  $8.881$
        & $1.049$
        &$ 0.306$
        &$ 20.644$\\
  & $\phi_{B1}$
        & $7.084$
        & $0.736$
        & $0.178$
        & $13.127$

   \\  \hline\hline

            & Mode
        &  $ \bar{B}_s^0 \rightarrow \bar{K}^{0}K^0$
        & $ \bar{B}_s^0 \rightarrow \pi^{0} \pi^{0}$
        &$ \bar{B}_s^0 \rightarrow \rho^0 K^{0} $
        & $ \bar{B}_s^0 \rightarrow \pi^{0} K^{*0}$
        \\  \hline
    & $\phi_{B1}+{\phi}_{B2}$
        & $19.566$
        & $0.525$
        & $0.085$
        & $0.113$\\
  & $\phi_{B1}$
        & $16.120$
        & $0.368$
        & $0.084$
        & $0.103$
          \\ \hline\hline       & Mode
        &$ \bar{B}_s^0 \rightarrow \omega K^{0}$
        &$ \bar{B}_s^0 \rightarrow K^{0} \phi$
        & $ \bar{B}_s^0 \rightarrow K^{*-} K^{+}$
        &$ \bar{B}_s^0 \rightarrow K^{-}K^{*+}$
        \\  \hline
    & $\phi_{B1}+{\phi}_{B2}$
        & $0.204$
        & $0.143$
        & $12.927$
        & $8.589$\\
  & $\phi_{B1}$
        & $0.127$
        & $0.150$
        & $8.853$
        & $5.474$
          \\ \hline\hline & Mode
        & $ \bar{B}_s^0 \rightarrow \bar{K}^{*0}K^{0}$
        &$ \bar{B}_s^0 \rightarrow \bar{K}^{0} K^{*0}$
        &$ \bar{B}_s^0 \rightarrow \pi^{-} K^{*+}$
        &$ \bar{B}_s^0 \rightarrow \rho^{-}K^{+}$
        \\  \hline
    & $\phi_{B1}+{\phi}_{B2}$
         &  $10.736$
        & $7.089$
        &$ 9.132$
        &$14.136$ \\
  & $\phi_{B1}$
        & $7.573$
        & $5.863$
        & $8.036$
        & $13.938$

   \\
          \hline\hline
  \end{tabular}
  \label{t11}
  \end{table}

\begin{table}[H]
\scriptsize
  \centering
  \caption{\small Comparison of theoretical predictions with experimental measurements for $\bar{B}^0_s \to PP, PV$ decays. The presented results include pQCD calculations at LO\cite{Ali:2007ff,Yan:2017nlj,Yan:2019nhf,Liu:2008rz} and NLO\cite{Yan:2017nlj,Yan:2019nhf,Liu:2008rz} accuracy, QCDF\cite{Ali:2007ff,Beneke:2003zv} predictions, and available experimental data from PDG\cite{ParticleDataGroup:2024cfk} and HFLAV\cite{HeavyFlavorAveragingGroupHFLAV:2024ctg}.}
  \begin{tabular}{ccccccccc}
  \hline\hline
    mode & \multicolumn{3}{c}{$\rm pQCD_{LO}$\cite{Ali:2007ff,Yan:2017nlj,Yan:2019nhf,Liu:2008rz}}%
         & \multicolumn{2}{c}{$\rm{pQCD_{\rm{NLO}}}$\cite{Yan:2017nlj,Yan:2019nhf,Liu:2008rz}}
         & $\rm{QCDF}$\cite{Ali:2007ff,Beneke:2003zv}

         & $\rm{PDG}$\cite{ParticleDataGroup:2024cfk}
            & $\rm{HFLAV}$\cite{HeavyFlavorAveragingGroupHFLAV:2024ctg}\\ \hline
   $ \bar{B}_s^0 \rightarrow \pi^{-} K^{+}$
  & $ 7.6_{-2.5}^{+3.3}$& 6.9&7.0&$5.70^{+2.3}_{-1.5}$ &$5.4^{+2.4}_{-1.5}$& $10.2$
   & $5.9\pm0.7$& $6.1^{+0.9}_{-0.8}$\\
    $ \bar{B}_s^0 \rightarrow \pi^{+} \pi^{-}$
  & $0.57_{-0.16}^{+0.18} $  &0.62 &0.70&$0.57_{-0.22}^{+0.24}$&$0.52_{-0.18}^{+0.21}$ & $0.024$
  & $0.72\pm0.1$ & $0.74_{-0.10}^{+0.12}$\\
    $ \bar{B}_s^0 \rightarrow \pi^{0} K^{0}$
  & $0.16_{-0.07}^{+0.12}$ & 0.18&0.16&$0.28^{+0.10}_{-0.08}$&$0.27^{+0.10}_{-0.08}$ & $0.49$
  & $-$ &$-$\\
   $ \bar{B}_s^0 \rightarrow K^{-} K^{+}$
  & $13.6_{-5.2}^{+8.6}$  &13.4 &11.8&$19.7^{+6.6}_{-5.7}$&$18.6^{+6.4}_{-5.3}$ & $22.7$
& $27.2\pm2.3$ & $27.4^{+3.2}_{-2.8}$\\
    $ \bar{B}_s^0 \rightarrow \bar{K}^{0} K^0 $
  & $15.6_{-6.0}^{+9.7} $  & 14.4&14.3&$20.2^{+7.3}_{-5.8}$ &$19.7^{+5.9}_{-3.8}$& $24.7$ & $17.6{\pm}3.1$& $17.4\pm3.1$ \\
   $ \bar{B}_s^0 \rightarrow \pi^{0} \pi^{0}$
  & $0.28_{-0.08}^{+0.09}$& 0.25&0.35&$0.29^{+0.12}_{-0.12}$&$0.21^{+0.10}_{-0.09}$ & $0.012$ & $<7.7$&$-$\\
    $  \bar{B}_s^0 \rightarrow\rho^0 K^{0}$
  & $  0.08_{-0.04}^{+0.07}$&0.10& $-$ & $-$ &  $0.34^{+0.12}_{-0.09}$ &$0.61$
   &  $-$ &  $-$\\
    $\bar{B}_s^0 \rightarrow \pi^{0} K^{*0}$
  & $0.07_{-0.02}^{+0.04} $ &0.08 &  $-$ & $-$ &$0.21_{-0.04}^{+0.07}$& $0.25$
  &  $-$ &  $-$ \\
    $ \bar{B}_s^0 \rightarrow  \omega  K^{0}$
  & $0.15_{-0.05}^{+0.08}$ &0.14 &  $-$ & $-$ &$0.65^{+0.22}_{-0.17}$&$0.51$
  & $-$&  $-$ \\
   $  \bar{B}_s^0 \rightarrow K^{0} \phi$
  & $0.16_{-0.05}^{+0.10}$&0.20 &  $-$ & $-$ &$024^{+0.05}_{-0.05}$  &$0.27$
&  $-$ &  $-$ \\
    $ \bar{B}_s^0 \rightarrow K^{*-} K^{+}$
  & $6.0_{-1.9}^{+2.5} $  &\multirow{2}{*}{9.03}& $-$  & $-$ & \multirow{2}{*}{$12.23^{+2.95}_{-3.41}$ } &$4.1^{+9.5}_{-3.2}$ &  \multirow{2}{*}{ $19{\pm}5$ }&  \multirow{2}{*}{ $18.6{\pm}4.70$ } \\
    $ \bar{B}_s^0 \rightarrow K^{-} K^{*+}$
  & $4.7_{-1.6}^{+2.7} $  && $-$  & $-$ &  &$5.5^{+15.1}_{-4.7}$ &  & \\
  
   $ \bar{B}_s^0 \rightarrow \bar{K}^{*0}K^{0}$
  & $ 7.3_{-2.1}^{+3.3}$& \multirow{2}{*}{9.95} &  $-$ & $-$ & \multirow{2}{*}{$14.39^{+3.54}_{-2.93}$ } &$3.9^{+10.6}_{-3.5}$
   &  \multirow{2}{*}{$20\pm6$}& \multirow{2}{*}{$19.8{\pm}5.7$}\\
    $ \bar{B}_s^0 \rightarrow \bar{K}^{0}K^{*0}$
  & $ 4.3_{-1.6}^{+2.3}$& &  $-$ & $-$ &  &$4.2^{+14.0}_{-4.0}$
   & &\\
    $ \bar{B}_s^0 \rightarrow \pi^{-} K^{*+}$
  & $ 7.6_{-2.3}^{+3.0}$  &6.32&  $-$ & $-$ &$3.96^{+1.41}_{-1.16}$&  $ 8.7$
   & $2.9\pm1.1$&$3.0\pm1.1$\\
    $  \bar{B}_s^0 \rightarrow \rho^{-}K^{+}$
  & $  17.8_{-5.89}^{+7.89} $ &18.6&  $-$ & $-$ &$15.9^{+6.5}_{-4.9}$&  $24.5$
 & $-$&  $-$ \\
  \hline\hline
  \end{tabular}
  \label{t05}
  \end{table}

\section{Conclusion \label{sec:summary}}

In this work, we revisit the decay processes $\bar{B}_s^0 \to PP$ and $PV$ using the pQCD method, incorporating the contribution of $\phi_{B2}$ in the wave function and the effects of higher-order terms in the final-state meson DAs. Using the minimum $\chi^2$ method, we determine the optimal value of $\omega_{B_s}$ and calculate the branching ratios and CP violation for these decay modes. Our results indicate that the inclusion of $\phi_{B2}$ has a noticeable impact on both the branching ratios and CP violation, providing a significant contribution to aligning the branching ratios of certain decay modes with experimental data. Thus, $\phi_{B2}$ should not be overlooked in the study of $B_s$ meson weak decays. 

Furthermore, while the contribution from higher-order terms in the final-state meson DAs is not as pronounced as that from $\phi_{B2}$, it still plays a role in improving the accuracy of the theoretical predictions. From an experimental perspective, future measurements will likely become increasingly precise. Theoretically, both corrections from $\phi_{B2}$ and other potential interaction mechanisms are crucial, underscoring the need for continued efforts on both the experimental and theoretical fronts.

\section*{Acknowledgments}
\addcontentsline{toc}{section}{Acknowledgements}
The work is supported by the National Natural Science Foundation of China (Grant Nos. 12275068, 12275067, 11705047), National Key R\&D Program of China (Grant No. 2023YFA1606000), Science and Technology R\&D Program Joint Fund Project of Henan Province (Grant No.225200810030),
Science and Technology Innovation Leading Talent Support Program of Henan Province, and Natural Science Foundation of Henan Province (Grant Nos. 222300420479, 242300420250).

\appendix

\section {The amplitudes for  \texorpdfstring{$\bar{B}_s^0 \to PP$ and $\bar{B}_s^0 \to PV$}{} }\label{sec:mode}
\subsection{Amplitude building blocks}
For conciseness, we adopt a compact notation for the hadronic WF components. In the framework of pQCD factorization, each hadron wave function is associated with a corresponding Sudakov factor, which takes the following specific form:
\begin{align}
  \phi_{P}^a&=\phi_{P}^a(x_2)e^{-S_{P}},\qquad \qquad \,\phi_{P^\prime}^a=\phi_{P^\prime}^a(x_3)e^{-S_{P^\prime}},\nonumber\\
  \phi_{P}^{p,t}&=r_P\phi_{P}^{p,t}(x_2)e^{-S_{P}},\qquad\quad
  \phi_{P^\prime}^{p,t}=r_{P}\phi_{P^\prime}^{p,t}(x_3)e^{-S_{P^\prime}},\nonumber\\
  \phi_{V}^{v}&=f_V^{\parallel}\phi_{V}^{v}(x_3)e^{-S_{V}},\qquad\quad
  \phi_{V}^{t,s}=r_Vf_V^{\perp}\phi_{V}^{t,s}(x_3)e^{-S_{V}}, \nonumber\\
  \phi_{B_1,B_2}&=\phi_{B_1,B_2}(x_1,b_1)e^{-S_{B_s}},\qquad {\cal C}=\frac{\pi C_F}{N_c^2}m_{B_s}^4f_{B_s}f_P.  
  \end{align}
  where $r_P=\mu_P/m_{B_s}$, $r_V=m_V/m_{B_s}$.


The explicit expressions for the decay amplitudes $\mathcal{A}_{a}^{LL}$, $\mathcal{A}_{a}^{LR}$, and $\mathcal{A}_{a}^{SP}$ corresponding to the factorizable emission diagram (a) in Fig. \ref{fig:ppchi}  are formulated as an illustrative case:
\begin{align}
\mathcal{A}_a^{LL}=&\int dx_{1}dx_{2}b_{1}db_{1}b_{2}db_{2}C_i(t_a)\alpha_s(t_a)S_t(x_2)H_{ab}(\alpha_g,\beta_a,b_1,b_2),\nonumber \\
& \{\phi_{B_1}[\phi_{M}^a(1+x_2)+ (\phi_{M}^{p}+\phi_{M}^{t}(\bar{x}_2-x_{2})]-\phi_{B_2}[\phi_{M}^a-(\phi_{M}^{p}+\phi_{M}^{t})x_2] \} \\
\mathcal{A}_a^{LR}=&-\mathcal{A}_a^{LL},\\
\mathcal{A}_a^{SP}=&2r_{M^\prime}\int dx_{1}dx_{2}b_{1}db_{1}b_{2}db_{2}C_i(t_a)\alpha_s(t_a)S_t(x_2)H_{ab}(\alpha_g,\beta_a,b_1,b_2)\nonumber \\
& \{\phi_{B_1}[\phi_{M}^a+ (\phi_{M}^{p}(2+x_2)-\phi_{M}^{t}x_2]-\phi_{B_2}[\phi_{M}^a-(\phi_{M}^{p}+\phi_{M}^{t})x_2] \}.
\end{align}
For a complete derivation of the  framework underlying all diagrammatic components, we direct readers to the comprehensive treatments in \cite{Yang:2020xal,Yang:2022ebu}.

\newpage
\subsection{Decay amplitudes for \texorpdfstring{$\bar{B}_s^0 \rightarrow PP$ }{}processes }

\begin{align}
 & {\cal A}(\bar B_s^0\to \pi^{-}K^{+}) \nonumber \\ 
=& \frac{G_{F}}{\sqrt{2}}V_{ub}V_{ud}^{*}[a_{1}{\cal A}_{ab}^{LL}(\pi^{-}K^{+})
+c_{1}A_{cd}^{LL}(\pi^{-}K^{+})]\nonumber\\
&-\frac{G_{F}}{\sqrt{2}}V_{tb}V_{td}^{*}[(a_{4}+a_{10}){\cal A}_{ab}^{LL}(\pi^{-}K^{+})
+(a_{6}+a_{8})A_{ab}^{SP}(\pi^{-}K^{+})\nonumber\\
&\qquad\qquad\quad+(c_3+c_{9})A_{cd}^{LL}(\pi^{-}K^{+})\nonumber\\
&\qquad\qquad\quad+(a_{4}-\frac{a_{10}}{2})A_{ef}^{LL}(\pi^{-}K^{+})
+(a_{6}-\frac{a_{8}}{2})A_{ef}^{SP}(\pi^{-}K^{+})\nonumber\\
&\qquad\qquad\quad+(c_3-\frac{c_{9}}{2})A_{gh}^{LL}(\pi^{-}K^{+})
+(c_5-\frac{c_{7}}{2})A_{gh}^{LR}(\pi^{-}K^{+})],  \\  &
   {\cal A}(\bar B_s^0\to \pi^{+}\pi^{-})=\sqrt{2 }{\cal A}(\bar B_s^0\to \pi^{0}\pi^{0})\nonumber\\
   &\qquad\qquad\quad\quad\quad=\frac{G_{F}}{\sqrt{2}}V_{ub}V_{us}^{*}\{c_{2}A_{gh}^{LL}\}-\frac{G_{F}}{\sqrt{2}}V_{tb}V_{ts}^{*}(2c_4+2c_6+\frac{c_{8}}{2}+\frac{c_{10}}{2})A_{gh}^{LL}(\pi^{+}\pi^{-}),\\
   & {\cal A}(\bar B_s^0\to \pi^{0}K^{0})
   \nonumber \\ =&
  \frac{G_{F}}{2}V_{ub}V_{ud}^{*}\{a_{2}{\cal A}_{ab}^{LL}(\pi^{0},K^{0})
   +{c_{2}}{\cal A}_{cd}^{LL}(\pi^{0},K^{0})\}\nonumber\\
   &-\frac{G_{F}}{2}V_{tb}V_{td}^{*}\{(-a_{4}-\frac{3a_{7}}{2}+\frac{3a_{9}}{2}+\frac{a_{10}}{2}){\cal A}_{ab}^{LL}(\pi^{0},K^{0})\nonumber\\
   &\qquad\qquad\quad+(-c_3+\frac{3c_{8}}{2}+\frac{c_9}{2}+\frac{3c_{10}}{2}){\cal A}_{cd}^{LL}(\pi^{0},K^{0})+(-a_{6}+\frac{a_{8}}{2}){\cal A}_{ab}^{SP}(\pi^{0},K^{0})\nonumber\\
   &\qquad\qquad\quad
   +(-a_{6}+\frac{a_{8}}{2}){\cal A}_{cd}^{SP}(\pi^{0},K^{0})+(-a_{4}+\frac{a_{10}}{2}){\cal A}_{ef}^{LL}(\pi^{0},K^{0})\nonumber\\
   &\qquad\qquad\quad+(-c_3+\frac{c_{9}}{2}){\cal A}_{gh}^{LL}(\pi^{0},K^{0})+(-c_5+\frac{c_{7}}{2}){\cal A}_{gh}^{LR}(\pi^{0},K^{0})\},\\  &
  {\cal A}(\bar B_s^0\to K^{-}K^{+})
   \nonumber \\ =&
  \frac{G_{F}}{\sqrt{2}}\{V_{ub}V_{us}^{*}\{a_{1}{\cal A}_{ab}^{LL}(K^{-},K^{+})
+{c_{1}}{\cal A}_{cd}^{LL}(K^{-},K^{+})+c_{2}{\cal A}_{gh}^{LL}(K^{-},K^{+})\}\nonumber \\
&- \frac{G_{F}}{\sqrt{2}}V_{tb}V_{ts}^{*}\{(a_{4}+a_{10}){\cal A}_{ab}^{LL}(K^{-},K^{+})
+(a_{6}+a_{8}){\cal A}_{ab}^{SP}(K^{-},K^{+}) \nonumber \\
 & \qquad\qquad\quad+(c_3+c_{9}){\cal A}_{cd}^{LL}(K^{-},K^{+})
+(c_5+c_{7}){\cal A}_{cd}^{LR}(K^{-},K^{+})\nonumber \\
& \qquad\qquad\quad +(a_{6}-\frac{a_{8}}{2}){\cal A}_{ef}^{SP}(K^{-},K^{+})+(c_{6}-\frac{c_{8}}{2}){\cal A}_{gh}^{SP}(K^{-},K^{+})\nonumber \\
&\qquad\qquad\quad+(c_3+c_{4}-\frac{c_{9}}{2}-\frac{c_{10}}{2}){\cal A}_{gh}^{LL}(K^{-},K^{+})+(c_5-\frac{c_{7}}{2}){\cal A}_{gh}^{LR}(K^{-},K^{+})\nonumber \\
 & \qquad\qquad\quad+(c_{4}+\frac{C_{10}}{2}){\cal A}_{gh}^{LL}(K^{+},K^{-})+(c_6+c_{8}){\cal A}_{gh}^{SP}(K^{+},K^{-})\},\\
& {\cal A}(\bar{B}_s^0 \to \bar{K}^0 K^0)
    \nonumber\\=&
  -\frac{G_F}{\sqrt{2}} V_{t b} V_{t s}^* \{(a_4-\frac{1}{2} a_{10})A_{ab}^{LL}(\bar{K}^0 K^0)+(a_6-\frac{1}{2} a_8)A_{ab}^{SP}( \bar{K}^0 K^0)\nonumber \\ & \qquad\qquad\quad +(C_3-\frac{1}{2} C_9)A_{cd}^{LL}(\bar{K}^0 K^0)+(C_5-\frac{1}{2} C_7)A_{cd}^{LR}(\bar{K}^0 K^0)\nonumber\\
  &\qquad\qquad\quad+(a_6-\frac{1}{2} a_8)A_{ef}^{S P}(\bar{K}^0 K^0)+(C_3-\frac{1}{2} C_9+C_4-\frac{1}{2} C_{10})A_{gh}^{LL}(\bar{K}^0 K^0)\nonumber\\
  &\qquad\qquad\quad+(C_6-\frac{1}{2} C_8)A_{gh}^{SP} (\bar{K}^0,K^0)+(C_5-\frac{1}{2} C_7)A_{gh}^{LR}(\bar{K}^0 K^0)\nonumber\\
  &\qquad\qquad\quad+(C_4-\frac{1}{2} C_{10})A_{gh}^{LL}(K^0,\bar{K}^0)+(C_6-\frac{1}{2} C_8)A_{gh}^{SP} (K^0,\bar{K}^0)\}.
   \end{align}
\subsection{Decay amplitudes for \texorpdfstring{$\bar{B}_s^0 \rightarrow PV$}{} processes}
 \begin{align} & 
{\cal A}(\bar{B}_s^0\rightarrow \rho^{0}K^{0})
   \nonumber\\=&
  \frac{G_{F}}{2}V_{ub}V_{ud}^{*}\{a_{2}A_{ab}^{LL}(\rho^{0},K^{0})
   +c_{2}A_{cd}^{LL}(\rho^{0},K^{0})\}\nonumber\\
   &-\frac{G_{F}}{2}V_{tb}V_{td}^{*}\{(-a_{4}+\frac{3a_{7}}{2}+\frac{3a_{9}}{2}+\frac{a_{10}}{2})A_{ab}^{LL}(\rho^{0},K^{0})\nonumber\\
   &\qquad\qquad\quad+(-c_3+\frac{c_9}{2}+\frac{3c_{10}}{2})A_{cd}^{LL}(\rho^{0},K^{0})+(\frac{3c_{8}}{2})A_{cd}^{SP}(\rho^{0},K^{0})\nonumber\\
   &\qquad\qquad\quad+(-c_{5}+\frac{c_7}{2})A_{cd}^{LR}(\rho^{0},K^{0})\nonumber\\
   &\qquad\qquad\quad+(-a_{4}+\frac{a_{10}}{2})A_{ef}^{LL}(\rho^{0},K^{0})
   +(-a_{6}+\frac{a_{8}}{2})A_{ef}^{SP}(\rho^{0},K^{0})\nonumber\\
    &\qquad\qquad\quad+(-c_3+\frac{c_{9}}{2})A_{gh}^{LL}(\rho^{0},K^{0})+(-c_5+\frac{c_{7}}{2})A_{gh}^{LR}(\rho^{0},K^{0})\},\\
    &
  {\cal A}(\bar B_s^0\to \pi^{0}K^{*0})
  \nonumber\\=&
  \frac{G_{F}}{2}V_{ub}V_{ud}^{*}\{a_{2}A_{ab}^{LL}(\pi^{0},K^{*0})
   +c_{2}A_{cd}^{LL}(\pi^{0},K^{*0})\}\nonumber\\
   &-\frac{G_{F}}{2}V_{tb}V_{td}^{*}\{(-a_{4}-\frac{3a_{7}}{2}+\frac{a_{10}}{2}+\frac{3a_{9}}{2})A_{ab}^{LL}(\pi^{0},K^{*0})\nonumber\\
   &\qquad\qquad\quad-(-a_{6}+\frac{a_{8}}{2})A_{ab}^{SP}(\pi^{0},K^{*0})+(-c_3+\frac{3c_{8}}{2}+\frac{c_9}{2}+\frac{3c_{10}}{2})A_{cd}^{LL}(\pi^{0},K^{*0})\nonumber\\
   &\qquad\qquad\quad+(-a_{4}+\frac{a_{10}}{2})A_{ef}^{LL}(\pi^{0},K^{*0})
   +(-a_{6}+\frac{a_{8}}{2})A_{ef}^{SP}(\pi^{0},K^{*0})\nonumber\\
   &\qquad\qquad\quad+(-c_3+\frac{c_{9}}{2})A_{gh}^{LL}(\pi^{0},K^{*0})-(-c_5+\frac{c_{7}}{2})A_{gh}^{LR}(\pi^{0},K^{*0})\},\\
   &
  {\cal A}(\bar B_s^0\rightarrow \omega K^{0} )
   \nonumber\\=&
  \frac{G_{F}}{2}V_{ub}V_{ud}^{*}\{a_{2}A_{ab}^{LL}(\omega ,K^{0} )
  +c_{2}A_{cd}^{LL}(\omega ,K^{0} )\}\nonumber\\
  &+\frac{G_{F}}{2}V_{tb}V_{td}^{*}\{(2a_{3}+a_4+2a_5+\frac{a_7}{2}+\frac{a_{9}}{2}-\frac{a_{10}}{2})A_{ab}^{LL}(\omega ,K^{0})\nonumber\\
  &\qquad\qquad\quad+(c_3+2c_4-\frac{c_9}{2}+\frac{c_{10}}{2})A_{cd}^{LL}(\omega ,K^{0} )-(2c_6+\frac{c_{8}}{2})A_{cd}^{SP}(\omega ,K^{0} ) \nonumber\\
  &\qquad\qquad\quad+(c_5-\frac{c_{7}}{2})A_{cd}^{LR}(\omega ,K^{0} )\nonumber \\
   &\qquad\qquad\quad+(a_{4}-\frac{a_{10}}{2})A_{ef}^{LL}(\omega ,K^{0} )+(a_{6}-\frac{a_{8}}{2})A_{ef}^{SP}(\omega ,K^{0} )\nonumber\\
   &\qquad\qquad\quad+(c_3-\frac{c_{9}}{2})A_{gh}^{LL}(\omega ,K^{0} )+(c_5-\frac{c_{7}}{2})A_{gh}^{LR}(\omega ,K^{0} ) \},\\
& {\cal A}(\bar B_s^0\rightarrow K^0 \phi)
\nonumber\\=&
-\frac{G_{F}}{\sqrt{2}}V_{tb}V_{td}^{*}\{(a_3+a_{5}-\frac{a_{7}}{2}-\frac{a_{9}}{2})A_{ab}^{LL}(\phi,K^{0})+(a_4-\frac{a_{10}}{2})A_{ab}^{LL}(K^{0},\phi)\nonumber\\
&\qquad\qquad\quad+(a_6-\frac{a_{8}}{2})A_{ab}^{SP}(K^{0},\phi)\nonumber\\
&\qquad\qquad\quad+(c_{4}-\frac{C_{10}}{2})A_{cd}^{LL}(\phi,K^{0})+(c_3-\frac{C_{9}}{2})A_{cd}^{LL}(K^{0},\phi)\nonumber \\
&\qquad\qquad\quad-(c_6-\frac{c_{8}}{2})A_{cd}^{SP}(\phi,K^{0})-(c_{5}-\frac{c_{7}}{2})A_{cd}^{LR}(K^{0},\phi)\nonumber\\
&\qquad\qquad\quad+(a_4-\frac{a_{10}}{2})A_{ef}^{LL}(K^{0},\phi)-(a_6-\frac{a_{8}}{2})A_{ef}^{SP}(K^{0},\phi)\nonumber\\
&\qquad\qquad\quad+(c_3-\frac{C_{9}}{2})A_{gh}^{LL}(K^{0},\phi)-(c_5-\frac{c_{7}}{2})A_{gh}^{LR}(K^{0},\phi)\},\\&
 {\cal A}(\bar B_s^0\rightarrow K^{*-}K^{+})
\nonumber\\=&
\frac{G_{F}}{\sqrt{2}}\{V_{ub}V_{us}^{*}\{a_{1}A_{ab}^{LL}(K^{*-},K^{+})+c_{1}A_{cd}^{LL}(K^{*-},K^{+})+a_{2}A_{ef}^{LL}(K^{+},K^{*-})+c_{2}A_{gh}^{LL}(K^{+},K^{*-})\}\nonumber\\
&-\frac{G_{F}}{\sqrt{2}}V_{tb}V_{ts}^{*}\{(a_{4}+a_{10})A_{ab}^{LL}(K^{*-},K^{+})+(c_3+c_{9})A_{cd}^{LL}(K^{*-},K^{+})+(c_5+c_{7})A_{cd}^{LR}(K^{*-},K^{+})\nonumber\\
&\qquad\qquad\quad+(a_3+a_{4}-a_5+\frac{a_7}{2}-\frac{a_9}{2}-\frac{a_{10}}{2})A_{ef}^{LL}(K^{*-},K^{+})\nonumber\\
&\qquad\qquad\quad+(a_{6}-\frac{a_{8}}{2})A_{ef}^{SP}(K^{*-},K^{+})+(a_3-a_5-a_7+a_9)A_{ef}^{LL}(K^{+},K^{*-})\nonumber\\
&\qquad\qquad\quad+(c_{5}-\frac{c_{7}}{2})A_{gh}^{LR}(K^{*-},K^{+})+(c_3+c_{4}-\frac{c_{9}}{2}-\frac{c_{10}}{2})A_{gh}^{LL}(K^{*-},K^{+})\nonumber\\
&\qquad\qquad\quad-(c_6-\frac{c_{8}}{2})A_{gh}^{SP}(K^{*-},K^+)
+(c_{4}+c_{10})A_{gh}^{LL}(K^{+},K^{*-})\nonumber\\
&\qquad\qquad\quad-(c_{6}+c_{8})A_{gh}^{SP}(K^{+},K^{*-})\},\\
& {\cal A}(\bar B_s^0\rightarrow K^{-}K^{*+})
 \nonumber\\=&
 \frac{G_{F}}{\sqrt{2}}\{V_{ub}V_{us}^{*}\{a_{1}A_{ab}^{LL}(K^{-},K^{*+})
+c_1A_{cd}^{LL}(K^{-},K^{*+})+a_{2}A_{ef}^{LL}(K^{*+},K^{-})
+c_{2}A_{gh}^{LL}(K^{*+},K^{-})\}\nonumber\\
&-\frac{G_{F}}{\sqrt{2}}V_{tb}V_{ts}^{*}\{(a_{4}+a_{10})A_{ab}^{LL}(K^{-},K^{*+})
-(a_{6}+a_{8})A_{ab}^{SP}(K^{-},K^{*+})-(c_{6}+c_{8})A_{gh}^{SP}(K^{*+},K^{-})\nonumber\\
&\qquad\qquad\quad+(c_3+c_{9})A_{cd}^{LL}(K^{-},K^{*+})-(c_5+c_{7})A_{cd}^{LL}(K^{-},K^{*+})\nonumber\\
&\qquad\qquad\quad+(a_3+a_{4}-a_5+\frac{a_7}{2}-\frac{a_9}{2}-\frac{a_{10}}{2})A_{ef}^{LL}(K^{-},K^{*+})\nonumber\\
&\qquad\qquad\quad+(a_3-a_5-a_7+a_9)A_{ef}^{LL}(K^{*+},K^{-})-(a_{6}-\frac{a_{8}}{2})A_{ef}^{SP}(K^{-},K^{*+})\nonumber\\
&\qquad\qquad\quad-(c_{5}-\frac{c_{7}}{2})A_{gh}^{LR}(K^{-},K^{*+})+(c_3+c_{4}-\frac{C_{9}}{2}-\frac{C_{10}}{2})A_{gh}^{LL}(K^{-},K^{*+})\nonumber\\
&\qquad\qquad\quad-(c_6-\frac{c_{8}}{2})A_{gh}^{SP}(K^{-},K^{*+})
+(c_{4}+c_{10})A_{gh}^{LL}(K^{*+},K^{-})\},\\
& {\cal A}(\bar B_s^0\rightarrow \bar{K}^{*0}K^{0})
\nonumber\\=&
-\frac{G_{F}}{\sqrt{2}}V_{tb}V_{ts}^{*}\{(a_{4}-\frac{a_{10}}{2})A_{ab}^{LL}(\bar{K}^{*0},K^0)+(c_3-\frac{c_{9}}{2})A_{cd}^{LL}(\bar{K}^{*0},K^0)\nonumber\\
&\qquad\qquad\quad+(c_5-\frac{c_{7}}{2})A_{cd}^{LR}(\bar{K}^{*0},K^0)+(a_{6}-\frac{a_{8}}{2})A_{ef}^{SP}(\bar{K}^{*0},K^0)\nonumber\\
&\qquad\qquad\quad+(a_3+a_{4}-a_5+\frac{a_7}{2}-\frac{a_9}{2}-\frac{a_{10}}{2})A_{ef}^{LL}(\bar{K}^{*0},K^0)\nonumber\\
&\qquad\qquad\quad+(c_3+c_{4}-\frac{c_{9}}{2}-\frac{c_{10}}{2})A_{gh}^{LL}(\bar{K}^{*0},K^0)+(c_4-\frac{c_{10}}{2})A_{gh}^{LL}(K^{0},\bar{K}^{*0})\nonumber\\
&\qquad\qquad\quad+(c_5-\frac{c_{7}}{2})A_{gh}^{LR}(\bar{K}^{*0},K^{0})-(c_{6}-\frac{c_{8}}{2})A_{gh}^{SP}(K^{0},\bar{K}^{*0})\nonumber\\
&\qquad\qquad\quad+(a_3-a_5+\frac{a_{7}}{2}-\frac{a_{9}}{2})A_{ef}^{LL}(K^{0},\bar{K}^{*0})-(c_{6}-\frac{c_{8}}{2})A_{gh}^{SP}(\bar{K}^{*0},K^{0})\},\\
& {\cal A}(\bar B_s^0\rightarrow \bar{K}^{0}K^{*0})
\nonumber\\=&
-\frac{G_{F}}{\sqrt{2}}V_{tb}V_{ts}^{*}\{(a_{4}-\frac{a_{10}}{2})A_{ab}^{LL}(\bar{K}^{0},K^{*0})+(a_{6}-\frac{a_{8}}{2})A_{ab}^{SP}(\bar{K}^{0},K^{*0})\nonumber\\
&\qquad\qquad\quad+(c_3-\frac{c_{9}}{2})A_{cd}^{LL}(\bar{K}^{0},K^{*0}-(c_5-\frac{c_{7}}{2})A_{cd}^{LR}(\bar{K}^{0},K^{*0})\nonumber\\
&\qquad\qquad\quad+(a_3+a_{4}-a_5+\frac{a_7}{2}-\frac{a_9}{2}-\frac{a_{10}}{2})A_{ef}^{LL}(\bar{K}^{0},K^{*0})\nonumber\\
&\qquad\qquad\quad-(a_{6}-\frac{a_{8}}{2})A_{ef}^{SP}(\bar{K}^{0},K^{*0})\nonumber\\
&\qquad\qquad\quad+(c_3+c_{4}-\frac{c_{9}}{2}-\frac{c_{10}}{2})A_{gh}^{LL}(\bar{K}^{0},K^{*0})+(c_4-\frac{c_{10}}{2})A_{gh}^{LL}(K^{*0},\bar{K}^{0})\nonumber\\
&\qquad\qquad\quad-(c_5-\frac{c_{7}}{2})A_{gh}^{LR}(\bar{K}^{0},K^{*0})-(c_{6}-\frac{c_{8}}{2})A_{gh}^{SP}(K^{*0},\bar{K}^{0})\nonumber\\
&\qquad\qquad\quad+(a_3-a_5+\frac{a_{7}}{2}-\frac{a_{9}}{2})A_{ef}^{LL}(K^{*0},\bar{K}^{0})-(c_{6}-\frac{c_{8}}{2})A_{gh}^{SP}(\bar{K}^{0},K^{*0})\},\\  
&{\cal A}(\bar B_s^0\rightarrow \pi^{-}K^{*+})
  \nonumber\\=&
  \frac{G_{F}}{\sqrt{2}}V_{ub}V_{ud}^{*}\{a_{1}A_{ab}^{LL}(\pi^{-},K^{*+})
   +c_{1}A_{cd}^{LL}(\pi^{-},K^{*+})\}\nonumber\\
   &-\frac{G_{F}}{\sqrt{2}}V_{tb}V_{td}^{*}\{(a_{4}+a_{10})A_{ab}^{LL}(\pi^{-},K^{*+})-(a_{6}+a_{8})A_{ab}^{SP}(\pi^{-},K^{*+})+(c_3+c_{9})A_{cd}^{LL}(\pi^{-},K^{*+})\nonumber\\
  &\qquad\qquad\quad+(a_{4}-\frac{a_{10}}{2})A_{ef}^{LL}(\pi^{-},K^{*+})
  -(a_{6}-\frac{a_{8}}{2})A_{ef}^{SP}(\pi^{-},K^{*+})\nonumber\\
   &\qquad\qquad\quad+(c_3-\frac{c_{9}}{2})A_{gh}^{LL}(\pi^{-},K^{*+})+(c_5-\frac{c_{7}}{2})A_{gh}^{LR}(\pi^{-},K^{*+})\},\\
   &{\cal A}(\bar B_s^0\rightarrow\rho^- K^+ )
  \nonumber\\=&
  \frac{G_{F}}{\sqrt{2}}V_{ub}V_{ud}^{*}\{a_{1}A_{ab}^{LL}(\rho^-,K^+)
   +c_{1}A_{cd}^{LL}(\rho^-,K^+)\}\nonumber\\
   &-\frac{G_{F}}{\sqrt{2}}V_{tb}V_{td}^{*}\{(a_{4}+a_{10})A_{ab}^{LL}(\rho^-,K^+)+(c_3+c_{9})A_{cd}^{LL}(\rho^-,K^+)+(c_5+c_{7})A_{cd}^{LR}(\rho^-,K^+)\nonumber\\
  &\qquad\qquad\quad+(a_{4}-\frac{a_{10}}{2})A_{ef}^{LL}(\rho^-,K^+)
  +(a_{6}-\frac{a_{8}}{2})A_{ef}^{SP}(\rho^-,K^+)\nonumber\\
   &\qquad\qquad\quad+(c_3-\frac{c_{9}}{2})A_{gh}^{LL}(\rho^-,K^+)+(c_5-\frac{c_{7}}{2})A_{gh}^{LR}(\rho^-,K^+)\}.
\end{align}

\medskip

\bibliographystyle{elsarticle-num-names}
\bibliography{paper}

\providecommand{\noopsort}[1]{}\providecommand{\singleletter}[1]{#1}%
\begin{thebibliography}{50}
\expandafter\ifx\csname natexlab\endcsname\relax\def\natexlab#1{#1}\fi
\providecommand{\url}[1]{\texttt{#1}}
\providecommand{\href}[2]{#2}
\providecommand{\path}[1]{#1}
\providecommand{\DOIprefix}{doi:}
\providecommand{\ArXivprefix}{arXiv:}
\providecommand{\URLprefix}{URL: }
\providecommand{\Pubmedprefix}{pmid:}
\providecommand{\doi}[1]{\href{http://dx.doi.org/#1}{\path{#1}}}
\providecommand{\Pubmed}[1]{\href{pmid:#1}{\path{#1}}}
\providecommand{\bibinfo}[2]{#2}
\ifx\xfnm\relax \def\xfnm[#1]{\unskip,\space#1}\fi
\bibitem[{Altmannshofer et~al.(2019)}]{Belle-II:2018jsg}
\bibinfo{author}{W.~Altmannshofer}, et~al. (\bibinfo{collaboration}{Belle-II
  Collaboration}),
\newblock \bibinfo{title}{{The Belle II Physics Book}},
\newblock \bibinfo{journal}{PTEP} \bibinfo{volume}{2019} (\bibinfo{year}{2019})
  \bibinfo{pages}{123C01}. \DOIprefix\doi{10.1093/ptep/ptz106}.
  \href{http://arxiv.org/abs/1808.10567}{{\tt arXiv:1808.10567}},
  \bibinfo{note}{[Erratum: PTEP 2020, 029201 (2020)]}.
\bibitem[{Aaij et~al.(2018)}]{LHCb:2018roe}
\bibinfo{author}{R.~Aaij}, et~al. (\bibinfo{collaboration}{LHCb
  Collaboration}),
\newblock \bibinfo{title}{{Physics case for an LHCb Upgrade II - Opportunities
  in flavour physics, and beyond, in the HL-LHC era}}  (\bibinfo{year}{2018}).
  \href{http://arxiv.org/abs/1808.08865}{{\tt arXiv:1808.08865}}.
\bibitem[{Abdallah et~al.(2024)}]{CEPCStudyGroup:2023quu}
\bibinfo{author}{W.~Abdallah}, et~al. (\bibinfo{collaboration}{CEPC Study
  Group}),
\newblock \bibinfo{title}{{CEPC Technical Design Report: Accelerator}},
\newblock \bibinfo{journal}{Radiat. Detect. Technol. Methods}
  \bibinfo{volume}{8} (\bibinfo{year}{2024}) \bibinfo{pages}{1--1105}.
  \DOIprefix\doi{10.1007/s41605-024-00463-y}.
  \href{http://arxiv.org/abs/2312.14363}{{\tt arXiv:2312.14363}},
  \bibinfo{note}{[Erratum: Radiat.Detect.Technol.Methods None, (2024)]}.
\bibitem[{Abada et~al.(2019)}]{FCC:2018byv}
\bibinfo{author}{A.~Abada}, et~al. (\bibinfo{collaboration}{FCC}),
\newblock \bibinfo{title}{{FCC Physics Opportunities}: {Future Circular
  Collider Conceptual Design Report Volume 1}},
\newblock \bibinfo{journal}{Eur. Phys. J. C} \bibinfo{volume}{79}
  (\bibinfo{year}{2019}) \bibinfo{pages}{474}.
  \DOIprefix\doi{10.1140/epjc/s10052-019-6904-3}.
\bibitem[{Navas et~al.(2024)}]{ParticleDataGroup:2024cfk}
\bibinfo{author}{S.~Navas}, et~al. (\bibinfo{collaboration}{Particle Data
  Group}),
\newblock \bibinfo{title}{{Review of particle physics}},
\newblock \bibinfo{journal}{Phys. Rev. D} \bibinfo{volume}{110}
  (\bibinfo{year}{2024}) \bibinfo{pages}{030001}.
  \DOIprefix\doi{10.1103/PhysRevD.110.030001}.
\bibitem[{Abdallah et~al.(2003)}]{DELPHI:2003pao}
\bibinfo{author}{J.~Abdallah}, et~al. (\bibinfo{collaboration}{DELPHI
  Collaboration}),
\newblock \bibinfo{title}{{A Measurement of the branching fractions of the $b$
  quark into charged and neutral $b$ hadrons}},
\newblock \bibinfo{journal}{Phys. Lett. B} \bibinfo{volume}{576}
  (\bibinfo{year}{2003}) \bibinfo{pages}{29--42}.
  \DOIprefix\doi{10.1016/j.physletb.2003.09.070}.
  \href{http://arxiv.org/abs/hep-ex/0311005}{{\tt arXiv:hep-ex/0311005}}.
\bibitem[{Aaij et~al.(2021)}]{LHCb:2021qbv}
\bibinfo{author}{R.~Aaij}, et~al. (\bibinfo{collaboration}{LHCb
  Collaboration}),
\newblock \bibinfo{title}{{Precise measurement of the~$f_s/f_d$ ratio of
  fragmentation fractions and of $B^0_s$ decay branching fractions}},
\newblock \bibinfo{journal}{Phys. Rev. D} \bibinfo{volume}{104}
  (\bibinfo{year}{2021}) \bibinfo{pages}{032005}.
  \DOIprefix\doi{10.1103/PhysRevD.104.032005}.
  \href{http://arxiv.org/abs/2103.06810}{{\tt arXiv:2103.06810}}.
\bibitem[{Aaltonen et~al.(2012)}]{CDF:2011who}
\bibinfo{author}{T.~Aaltonen}, et~al. (\bibinfo{collaboration}{CDF
  Collaboration}),
\newblock \bibinfo{title}{{Evidence for the charmless annihilation decay mode
  $B^0_s \to \pi^+\pi^-$}},
\newblock \bibinfo{journal}{Phys. Rev. Lett.} \bibinfo{volume}{108}
  (\bibinfo{year}{2012}) \bibinfo{pages}{211803}.
  \DOIprefix\doi{10.1103/PhysRevLett.108.211803}.
  \href{http://arxiv.org/abs/1111.0485}{{\tt arXiv:1111.0485}}.
\bibitem[{Aaltonen et~al.(2009)}]{CDF:2008llm}
\bibinfo{author}{T.~Aaltonen}, et~al. (\bibinfo{collaboration}{CDF
  Collaboration}),
\newblock \bibinfo{title}{{Observation of New Charmless Decays of Bottom
  Hadrons}},
\newblock \bibinfo{journal}{Phys. Rev. Lett.} \bibinfo{volume}{103}
  (\bibinfo{year}{2009}) \bibinfo{pages}{031801}.
  \DOIprefix\doi{10.1103/PhysRevLett.103.031801}.
  \href{http://arxiv.org/abs/0812.4271}{{\tt arXiv:0812.4271}}.
\bibitem[{Aaltonen et~al.(2011)}]{CDF:2011ubb}
\bibinfo{author}{T.~Aaltonen}, et~al. (\bibinfo{collaboration}{CDF
  Collaboration}),
\newblock \bibinfo{title}{{Measurements of Direct CP Violating Asymmetries in
  Charmless Decays of Strange Bottom Mesons and Bottom Baryons}},
\newblock \bibinfo{journal}{Phys. Rev. Lett.} \bibinfo{volume}{106}
  (\bibinfo{year}{2011}) \bibinfo{pages}{181802}.
  \DOIprefix\doi{10.1103/PhysRevLett.106.181802}.
  \href{http://arxiv.org/abs/1103.5762}{{\tt arXiv:1103.5762}}.
\bibitem[{Aaltonen et~al.(2014)}]{CDF:2014pzb}
\bibinfo{author}{T.~A. Aaltonen}, et~al. (\bibinfo{collaboration}{CDF
  Collaboration}),
\newblock \bibinfo{title}{{Measurements of Direct CP -Violating Asymmetries in
  Charmless Decays of Bottom Baryons}},
\newblock \bibinfo{journal}{Phys. Rev. Lett.} \bibinfo{volume}{113}
  (\bibinfo{year}{2014}) \bibinfo{pages}{242001}.
  \DOIprefix\doi{10.1103/PhysRevLett.113.242001}.
  \href{http://arxiv.org/abs/1403.5586}{{\tt arXiv:1403.5586}}.
\bibitem[{Peng et~al.(2010)}]{Belle:2010yix}
\bibinfo{author}{C.~C. Peng}, et~al. (\bibinfo{collaboration}{Belle
  Collaboration}),
\newblock \bibinfo{title}{{Search for $B_{s}^{0} \to hh$ Decays at the
  $\Upsilon(5S)$ Resonance}},
\newblock \bibinfo{journal}{Phys. Rev. D} \bibinfo{volume}{82}
  (\bibinfo{year}{2010}) \bibinfo{pages}{072007}.
  \DOIprefix\doi{10.1103/PhysRevD.82.072007}.
  \href{http://arxiv.org/abs/1006.5115}{{\tt arXiv:1006.5115}}.
\bibitem[{Pal et~al.(2016)}]{Belle:2015gho}
\bibinfo{author}{B.~Pal}, et~al. (\bibinfo{collaboration}{Belle
  Collaboration}),
\newblock \bibinfo{title}{{Observation of the decay $B_s^0\rightarrow
  K^0\overline{K}^0$}},
\newblock \bibinfo{journal}{Phys. Rev. Lett.} \bibinfo{volume}{116}
  (\bibinfo{year}{2016}) \bibinfo{pages}{161801}.
  \DOIprefix\doi{10.1103/PhysRevLett.116.161801}.
  \href{http://arxiv.org/abs/1512.02145}{{\tt arXiv:1512.02145}}.
\bibitem[{Aaij et~al.(2017)}]{LHCb:2016inp}
\bibinfo{author}{R.~Aaij}, et~al. (\bibinfo{collaboration}{LHCb
  Collaboration}),
\newblock \bibinfo{title}{{Observation of the annihilation decay mode $B^{0}\to
  K^{+}K^{-}$}},
\newblock \bibinfo{journal}{Phys. Rev. Lett.} \bibinfo{volume}{118}
  (\bibinfo{year}{2017}) \bibinfo{pages}{081801}.
  \DOIprefix\doi{10.1103/PhysRevLett.118.081801}.
  \href{http://arxiv.org/abs/1610.08288}{{\tt arXiv:1610.08288}}.
\bibitem[{Aaij et~al.(2012)}]{LHCb:2012ihl}
\bibinfo{author}{R.~Aaij}, et~al. (\bibinfo{collaboration}{LHCb
  Collaboration}),
\newblock \bibinfo{title}{{Measurement of $b$-hadron branching fractions for
  two-body decays into charmless charged hadrons}},
\newblock \bibinfo{journal}{JHEP} \bibinfo{volume}{10} (\bibinfo{year}{2012})
  \bibinfo{pages}{037}. \DOIprefix\doi{10.1007/JHEP10(2012)037}.
  \href{http://arxiv.org/abs/1206.2794}{{\tt arXiv:1206.2794}}.
\bibitem[{Aaij et~al.(2020)}]{LHCb:2020wrt}
\bibinfo{author}{R.~Aaij}, et~al. (\bibinfo{collaboration}{LHCb
  Collaboration}),
\newblock \bibinfo{title}{{Measurement of the branching fraction of the decay
  $B_s^0\to K_S^0 K_S^0$}},
\newblock \bibinfo{journal}{Phys. Rev. D} \bibinfo{volume}{102}
  (\bibinfo{year}{2020}) \bibinfo{pages}{012011}.
  \DOIprefix\doi{10.1103/PhysRevD.102.012011}.
  \href{http://arxiv.org/abs/2002.08229}{{\tt arXiv:2002.08229}}.
\bibitem[{Aaij et~al.(2019)}]{LHCb:2019vww}
\bibinfo{author}{R.~Aaij}, et~al. (\bibinfo{collaboration}{LHCb
  Collaboration}),
\newblock \bibinfo{title}{{Amplitude analysis of $B^{0}_{s} \rightarrow
  K^{0}_{\textrm{S}} K^{\pm}\pi^{\mp}$ decays}},
\newblock \bibinfo{journal}{JHEP} \bibinfo{volume}{06} (\bibinfo{year}{2019})
  \bibinfo{pages}{114}. \DOIprefix\doi{10.1007/JHEP06(2019)114}.
  \href{http://arxiv.org/abs/1902.07955}{{\tt arXiv:1902.07955}}.
\bibitem[{Aaij et~al.(2014)}]{LHCb:2014lcy}
\bibinfo{author}{R.~Aaij}, et~al. (\bibinfo{collaboration}{LHCb
  Collaboration}),
\newblock \bibinfo{title}{{Observation of $B^0_s \to K^{*\pm}K^\mp$ and
  evidence for $B^0_s \to K^{*-}\pi^+$ decays}},
\newblock \bibinfo{journal}{New J. Phys.} \bibinfo{volume}{16}
  (\bibinfo{year}{2014}) \bibinfo{pages}{123001}.
  \DOIprefix\doi{10.1088/1367-2630/16/12/123001}.
  \href{http://arxiv.org/abs/1407.7704}{{\tt arXiv:1407.7704}}.
\bibitem[{Aaij et~al.(2018)}]{LHCb:2018pff}
\bibinfo{author}{R.~Aaij}, et~al. (\bibinfo{collaboration}{LHCb
  Collaboration}),
\newblock \bibinfo{title}{{Measurement of $C\!P$ asymmetries in two-body
  $B_{(s)}^{0}$-meson decays to charged pions and kaons}},
\newblock \bibinfo{journal}{Phys. Rev. D} \bibinfo{volume}{98}
  (\bibinfo{year}{2018}) \bibinfo{pages}{032004}.
  \DOIprefix\doi{10.1103/PhysRevD.98.032004}.
  \href{http://arxiv.org/abs/1805.06759}{{\tt arXiv:1805.06759}}.
\bibitem[{Aaij et~al.(2021)}]{LHCb:2020byh}
\bibinfo{author}{R.~Aaij}, et~al. (\bibinfo{collaboration}{LHCb
  Collaboration}),
\newblock \bibinfo{title}{{Observation of $CP$ violation in two-body $
  {B}_{(s)}^0 $-meson decays to charged pions and kaons}},
\newblock \bibinfo{journal}{JHEP} \bibinfo{volume}{03} (\bibinfo{year}{2021})
  \bibinfo{pages}{075}. \DOIprefix\doi{10.1007/JHEP03(2021)075}.
  \href{http://arxiv.org/abs/2012.05319}{{\tt arXiv:2012.05319}}.
\bibitem[{Aaij et~al.(2013)}]{LHCb:2013clb}
\bibinfo{author}{R.~Aaij}, et~al. (\bibinfo{collaboration}{LHCb
  Collaboration}),
\newblock \bibinfo{title}{{First measurement of time-dependent $C\!P$ violation
  in $B^0_s \to K^+K^-$ decays}},
\newblock \bibinfo{journal}{JHEP} \bibinfo{volume}{10} (\bibinfo{year}{2013})
  \bibinfo{pages}{183}. \DOIprefix\doi{10.1007/JHEP10(2013)183}.
  \href{http://arxiv.org/abs/1308.1428}{{\tt arXiv:1308.1428}}.
\bibitem[{Banerjee et~al.(2024)}]{HeavyFlavorAveragingGroupHFLAV:2024ctg}
\bibinfo{author}{S.~Banerjee}, et~al. (\bibinfo{collaboration}{Heavy Flavor
  Averaging Group (HFLAV)}),
\newblock \bibinfo{title}{{Averages of $b$-hadron, $c$-hadron, and
  $\tau$-lepton properties as of 2023}}  (\bibinfo{year}{2024}).
  \href{http://arxiv.org/abs/2411.18639}{{\tt arXiv:2411.18639}}.
\bibitem[{Beneke and Neubert(2003)}]{Beneke:2003zv}
\bibinfo{author}{M.~Beneke}, \bibinfo{author}{M.~Neubert},
\newblock \bibinfo{title}{{QCD factorization for $B \to PP$ and $B \to PV$
  decays}},
\newblock \bibinfo{journal}{Nucl. Phys. B} \bibinfo{volume}{675}
  (\bibinfo{year}{2003}) \bibinfo{pages}{333--415}.
  \DOIprefix\doi{10.1016/j.nuclphysb.2003.09.026}.
  \href{http://arxiv.org/abs/hep-ph/0308039}{{\tt arXiv:hep-ph/0308039}}.
\bibitem[{Sun et~al.(2003)Sun, Zhu, and Du}]{Sun:2002rn}
\bibinfo{author}{J.-f. Sun}, \bibinfo{author}{G.-h. Zhu},
  \bibinfo{author}{D.-s. Du},
\newblock \bibinfo{title}{{Phenomenological analysis of charmless decays $B_s
  \to PP, PV$, with QCD factorization}},
\newblock \bibinfo{journal}{Phys. Rev. D} \bibinfo{volume}{68}
  (\bibinfo{year}{2003}) \bibinfo{pages}{054003}.
  \DOIprefix\doi{10.1103/PhysRevD.68.054003}.
  \href{http://arxiv.org/abs/hep-ph/0211154}{{\tt arXiv:hep-ph/0211154}}.
\bibitem[{Cheng and Chua(2009)}]{Cheng:2009mu}
\bibinfo{author}{H.-Y. Cheng}, \bibinfo{author}{C.-K. Chua},
\newblock \bibinfo{title}{{{QCD} Factorization for Charmless Hadronic $B_s$
  Decays Revisited}},
\newblock \bibinfo{journal}{Phys. Rev. D} \bibinfo{volume}{80}
  (\bibinfo{year}{2009}) \bibinfo{pages}{114026}.
  \DOIprefix\doi{10.1103/PhysRevD.80.114026}.
  \href{http://arxiv.org/abs/0910.5237}{{\tt arXiv:0910.5237}}.
\bibitem[{Chang et~al.(2015)Chang, Sun, Yang, and Li}]{Chang:2014yma}
\bibinfo{author}{Q.~Chang}, \bibinfo{author}{J.~Sun},
  \bibinfo{author}{Y.~Yang}, \bibinfo{author}{X.~Li},
\newblock \bibinfo{title}{{A combined fit on the annihilation corrections in
  $B_{u,d,s} \to PP$ decays within QCDF}},
\newblock \bibinfo{journal}{Phys. Lett. B} \bibinfo{volume}{740}
  (\bibinfo{year}{2015}) \bibinfo{pages}{56--60}.
  \DOIprefix\doi{10.1016/j.physletb.2014.11.027}.
  \href{http://arxiv.org/abs/1409.2995}{{\tt arXiv:1409.2995}}.
\bibitem[{Williamson and Zupan(2006)}]{Williamson:2006hb}
\bibinfo{author}{A.~R. Williamson}, \bibinfo{author}{J.~Zupan},
\newblock \bibinfo{title}{{Two body B decays with isosinglet final states in
  SCET}},
\newblock \bibinfo{journal}{Phys. Rev. D} \bibinfo{volume}{74}
  (\bibinfo{year}{2006}) \bibinfo{pages}{014003}.
  \DOIprefix\doi{10.1103/PhysRevD.74.014003}.
  \href{http://arxiv.org/abs/hep-ph/0601214}{{\tt arXiv:hep-ph/0601214}},
  \bibinfo{note}{[Erratum: Phys.Rev.D 74, 03901 (2006)]}.
\bibitem[{Li et~al.(2004)Li, Lu, Xiao, and Yu}]{Li:2004ep}
\bibinfo{author}{Y.~Li}, \bibinfo{author}{C.-D. Lu}, \bibinfo{author}{Z.-J.
  Xiao}, \bibinfo{author}{X.-Q. Yu},
\newblock \bibinfo{title}{{Branching ratio and CP asymmetry of $B_s\to \pi^+
  \pi^-$ decays in the perturbative QCD approach}},
\newblock \bibinfo{journal}{Phys. Rev. D} \bibinfo{volume}{70}
  (\bibinfo{year}{2004}) \bibinfo{pages}{034009}.
  \DOIprefix\doi{10.1103/PhysRevD.70.034009}.
  \href{http://arxiv.org/abs/hep-ph/0404028}{{\tt arXiv:hep-ph/0404028}}.
\bibitem[{Ali et~al.(2007)Ali, Kramer, Li, Lu, Shen, Wang, and
  Wang}]{Ali:2007ff}
\bibinfo{author}{A.~Ali}, \bibinfo{author}{G.~Kramer}, \bibinfo{author}{Y.~Li},
  \bibinfo{author}{C.-D. Lu}, \bibinfo{author}{Y.-L. Shen},
  \bibinfo{author}{W.~Wang}, \bibinfo{author}{Y.-M. Wang},
\newblock \bibinfo{title}{{Charmless non-leptonic $B_s$ decays to $PP$, $PV$
  and $VV$ final states in the pQCD approach}},
\newblock \bibinfo{journal}{Phys. Rev. D} \bibinfo{volume}{76}
  (\bibinfo{year}{2007}) \bibinfo{pages}{074018}.
  \DOIprefix\doi{10.1103/PhysRevD.76.074018}.
  \href{http://arxiv.org/abs/hep-ph/0703162}{{\tt arXiv:hep-ph/0703162}}.
\bibitem[{Xiao et~al.(2012)Xiao, Wang, and Fan}]{Xiao:2011tx}
\bibinfo{author}{Z.-J. Xiao}, \bibinfo{author}{W.-F. Wang},
  \bibinfo{author}{Y.-y. Fan},
\newblock \bibinfo{title}{{Revisiting the pure annihilation decays $B_s\to
  \pi^+ \pi^-$ and $B^0 \to K^+ K^-$: the data and the pQCD predictions}},
\newblock \bibinfo{journal}{Phys. Rev. D} \bibinfo{volume}{85}
  (\bibinfo{year}{2012}) \bibinfo{pages}{094003}.
  \DOIprefix\doi{10.1103/PhysRevD.85.094003}.
  \href{http://arxiv.org/abs/1111.6264}{{\tt arXiv:1111.6264}}.
\bibitem[{Wang et~al.(2014)Wang, Lin, Sun, Ji, Cheng, and Xiao}]{Wang:2014mua}
\bibinfo{author}{J.-J. Wang}, \bibinfo{author}{D.-T. Lin},
  \bibinfo{author}{W.~Sun}, \bibinfo{author}{Z.-J. Ji},
  \bibinfo{author}{S.~Cheng}, \bibinfo{author}{Z.-J. Xiao},
\newblock \bibinfo{title}{{$\bar{B}^0_s \to K\pi,KK$ decays and effects of the
  next-to-leading order contributions}},
\newblock \bibinfo{journal}{Phys. Rev. D} \bibinfo{volume}{89}
  (\bibinfo{year}{2014}) \bibinfo{pages}{074046}.
  \DOIprefix\doi{10.1103/PhysRevD.89.074046}.
  \href{http://arxiv.org/abs/1402.6912}{{\tt arXiv:1402.6912}}.
\bibitem[{Yan et~al.(2019)Yan, Liu, and Xiao}]{Yan:2019nhf}
\bibinfo{author}{D.-C. Yan}, \bibinfo{author}{X.~Liu}, \bibinfo{author}{Z.-J.
  Xiao},
\newblock \bibinfo{title}{{Anatomy of $B_s \to PP $ decays and effects of the
  next-to-leading order contributions in the perturbative QCD approach}},
\newblock \bibinfo{journal}{Nucl. Phys. B} \bibinfo{volume}{946}
  (\bibinfo{year}{2019}) \bibinfo{pages}{114705}.
  \DOIprefix\doi{10.1016/j.nuclphysb.2019.114705}.
  \href{http://arxiv.org/abs/1906.01442}{{\tt arXiv:1906.01442}}.
\bibitem[{Yan et~al.(2018)Yan, Yang, Liu, and Xiao}]{Yan:2017nlj}
\bibinfo{author}{D.-C. Yan}, \bibinfo{author}{P.~Yang},
  \bibinfo{author}{X.~Liu}, \bibinfo{author}{Z.-J. Xiao},
\newblock \bibinfo{title}{{Anatomy of $B_s \to PV $ decays and effects of
  next-to-leading order contributions in the perturbative QCD factorization
  approach}},
\newblock \bibinfo{journal}{Nucl. Phys. B} \bibinfo{volume}{931}
  (\bibinfo{year}{2018}) \bibinfo{pages}{79--104}.
  \DOIprefix\doi{10.1016/j.nuclphysb.2018.04.007}.
  \href{http://arxiv.org/abs/1707.06043}{{\tt arXiv:1707.06043}}.
\bibitem[{Cheng and Oh(2011)}]{Cheng:2011qh}
\bibinfo{author}{H.-Y. Cheng}, \bibinfo{author}{S.~Oh},
\newblock \bibinfo{title}{{Flavor SU(3) symmetry and QCD factorization in $B
  \to PP$ and $PV$ decays}},
\newblock \bibinfo{journal}{JHEP} \bibinfo{volume}{09} (\bibinfo{year}{2011})
  \bibinfo{pages}{024}. \DOIprefix\doi{10.1007/JHEP09(2011)024}.
  \href{http://arxiv.org/abs/1104.4144}{{\tt arXiv:1104.4144}}.
\bibitem[{Cheng et~al.(2015)Cheng, Chiang, and Kuo}]{Cheng:2014rfa}
\bibinfo{author}{H.-Y. Cheng}, \bibinfo{author}{C.-W. Chiang},
  \bibinfo{author}{A.-L. Kuo},
\newblock \bibinfo{title}{{Updating B\textrightarrow{}PP,VP decays in the
  framework of flavor symmetry}},
\newblock \bibinfo{journal}{Phys. Rev. D} \bibinfo{volume}{91}
  (\bibinfo{year}{2015}) \bibinfo{pages}{014011}.
  \DOIprefix\doi{10.1103/PhysRevD.91.014011}.
  \href{http://arxiv.org/abs/1409.5026}{{\tt arXiv:1409.5026}}.
\bibitem[{Yang et~al.(2021)Yang, Lang, Zhao, Huang, and Sun}]{Yang:2020xal}
\bibinfo{author}{Y.~Yang}, \bibinfo{author}{L.~Lang},
  \bibinfo{author}{X.~Zhao}, \bibinfo{author}{J.~Huang},
  \bibinfo{author}{J.~Sun},
\newblock \bibinfo{title}{{Reinvestigating the $B$ ${\to}$ $PP$ decays by
  including the contributions from ${\phi}_{B2}$}},
\newblock \bibinfo{journal}{Phys. Rev. D} \bibinfo{volume}{103}
  (\bibinfo{year}{2021}) \bibinfo{pages}{056006}.
  \DOIprefix\doi{10.1103/PhysRevD.103.056006}.
  \href{http://arxiv.org/abs/2012.10581}{{\tt arXiv:2012.10581}}.
\bibitem[{Yang et~al.(2022)Yang, Zhao, Lang, Huang, and Sun}]{Yang:2022ebu}
\bibinfo{author}{Y.~Yang}, \bibinfo{author}{X.~Zhao},
  \bibinfo{author}{L.~Lang}, \bibinfo{author}{J.~Huang},
  \bibinfo{author}{J.~Sun},
\newblock \bibinfo{title}{{Reinvestigating $B \to PV$ decays by including
  contributions from ${\phi}_{B2}$ with the perturbative QCD approach *}},
\newblock \bibinfo{journal}{Chin. Phys. C} \bibinfo{volume}{46}
  (\bibinfo{year}{2022}) \bibinfo{pages}{083103}.
  \DOIprefix\doi{10.1088/1674-1137/ac6573}.
  \href{http://arxiv.org/abs/2201.12834}{{\tt arXiv:2201.12834}}.
\bibitem[{Ball et~al.(2006)Ball, Braun, and Lenz}]{Ball:2006wn}
\bibinfo{author}{P.~Ball}, \bibinfo{author}{V.~M. Braun},
  \bibinfo{author}{A.~Lenz},
\newblock \bibinfo{title}{{Higher-twist distribution amplitudes of the $K$
  meson in QCD}},
\newblock \bibinfo{journal}{JHEP} \bibinfo{volume}{05} (\bibinfo{year}{2006})
  \bibinfo{pages}{004}. \DOIprefix\doi{10.1088/1126-6708/2006/05/004}.
  \href{http://arxiv.org/abs/hep-ph/0603063}{{\tt arXiv:hep-ph/0603063}}.
\bibitem[{Huang and Wu(2005)}]{Huang:2004hw}
\bibinfo{author}{T.~Huang}, \bibinfo{author}{X.-G. Wu},
\newblock \bibinfo{title}{{Consistent calculation of the $B \to \pi$ transition
  form-factor in the whole physical region}},
\newblock \bibinfo{journal}{Phys. Rev. D} \bibinfo{volume}{71}
  (\bibinfo{year}{2005}) \bibinfo{pages}{034018}.
  \DOIprefix\doi{10.1103/PhysRevD.71.034018}.
  \href{http://arxiv.org/abs/hep-ph/0412417}{{\tt arXiv:hep-ph/0412417}}.
\bibitem[{Kurimoto(2006)}]{Kurimoto:2006iv}
\bibinfo{author}{T.~Kurimoto},
\newblock \bibinfo{title}{{Uncertainty in the leading order PQCD calculations
  of B meson decays}},
\newblock \bibinfo{journal}{Phys. Rev. D} \bibinfo{volume}{74}
  (\bibinfo{year}{2006}) \bibinfo{pages}{014027}.
  \DOIprefix\doi{10.1103/PhysRevD.74.014027}.
  \href{http://arxiv.org/abs/hep-ph/0605112}{{\tt arXiv:hep-ph/0605112}}.
\bibitem[{Cheng et~al.(2014)Cheng, Fan, Yu, L\"u, and Xiao}]{Cheng:2014fwa}
\bibinfo{author}{S.~Cheng}, \bibinfo{author}{Y.-Y. Fan},
  \bibinfo{author}{X.~Yu}, \bibinfo{author}{C.-D. L\"u}, \bibinfo{author}{Z.-J.
  Xiao},
\newblock \bibinfo{title}{{The NLO twist-3 contributions to $B \to \pi$ form
  factors in $k_{T}$ factorization}},
\newblock \bibinfo{journal}{Phys. Rev. D} \bibinfo{volume}{89}
  (\bibinfo{year}{2014}) \bibinfo{pages}{094004}.
  \DOIprefix\doi{10.1103/PhysRevD.89.094004}.
  \href{http://arxiv.org/abs/1402.5501}{{\tt arXiv:1402.5501}}.
\bibitem[{Lu and Yang(2002)}]{Lu:2000hj}
\bibinfo{author}{C.-D. Lu}, \bibinfo{author}{M.-Z. Yang},
\newblock \bibinfo{title}{{$B \to \pi \rho, \pi \omega$ decays in perturbative
  QCD approach}},
\newblock \bibinfo{journal}{Eur. Phys. J. C} \bibinfo{volume}{23}
  (\bibinfo{year}{2002}) \bibinfo{pages}{275--287}.
  \DOIprefix\doi{10.1007/s100520100878}.
  \href{http://arxiv.org/abs/hep-ph/0011238}{{\tt arXiv:hep-ph/0011238}}.
\bibitem[{Descotes-Genon and Sachrajda(2002)}]{Descotes-Genon:2001rya}
\bibinfo{author}{S.~Descotes-Genon}, \bibinfo{author}{C.~T. Sachrajda},
\newblock \bibinfo{title}{{Sudakov effects in $B \to \pi \ell \nu_{\ell}$
  form-factors}},
\newblock \bibinfo{journal}{Nucl. Phys. B} \bibinfo{volume}{625}
  (\bibinfo{year}{2002}) \bibinfo{pages}{239--278}.
  \DOIprefix\doi{10.1016/S0550-3213(02)00017-2}.
  \href{http://arxiv.org/abs/hep-ph/0109260}{{\tt arXiv:hep-ph/0109260}}.
\bibitem[{Wei and Yang(2002)}]{Wei:2002iu}
\bibinfo{author}{Z.-T. Wei}, \bibinfo{author}{M.-Z. Yang},
\newblock \bibinfo{title}{{The Systematic study of $B \to \pi$ form-factors in
  pQCD approach and its reliability}},
\newblock \bibinfo{journal}{Nucl. Phys. B} \bibinfo{volume}{642}
  (\bibinfo{year}{2002}) \bibinfo{pages}{263--289}.
  \DOIprefix\doi{10.1016/S0550-3213(02)00623-5}.
  \href{http://arxiv.org/abs/hep-ph/0202018}{{\tt arXiv:hep-ph/0202018}}.
\bibitem[{Keum et~al.(2001)Keum, Li, and Sanda}]{Keum:2000wi}
\bibinfo{author}{Y.~Y. Keum}, \bibinfo{author}{H.-N. Li},
  \bibinfo{author}{A.~I. Sanda},
\newblock \bibinfo{title}{{Penguin enhancement and $B \to K \pi$ decays in
  perturbative QCD}},
\newblock \bibinfo{journal}{Phys. Rev. D} \bibinfo{volume}{63}
  (\bibinfo{year}{2001}) \bibinfo{pages}{054008}.
  \DOIprefix\doi{10.1103/PhysRevD.63.054008}.
  \href{http://arxiv.org/abs/hep-ph/0004173}{{\tt arXiv:hep-ph/0004173}}.
\bibitem[{Ball(1999)}]{Ball:1998je}
\bibinfo{author}{P.~Ball},
\newblock \bibinfo{title}{{Theoretical update of pseudoscalar meson
  distribution amplitudes of higher twist: The Nonsinglet case}},
\newblock \bibinfo{journal}{JHEP} \bibinfo{volume}{01} (\bibinfo{year}{1999})
  \bibinfo{pages}{010}. \DOIprefix\doi{10.1088/1126-6708/1999/01/010}.
  \href{http://arxiv.org/abs/hep-ph/9812375}{{\tt arXiv:hep-ph/9812375}}.
\bibitem[{Ball and Jones(2007)}]{Ball:2007rt}
\bibinfo{author}{P.~Ball}, \bibinfo{author}{G.~W. Jones},
\newblock \bibinfo{title}{{Twist-3 distribution amplitudes of $K^*$ and $\phi$
  mesons}},
\newblock \bibinfo{journal}{JHEP} \bibinfo{volume}{03} (\bibinfo{year}{2007})
  \bibinfo{pages}{069}. \DOIprefix\doi{10.1088/1126-6708/2007/03/069}.
  \href{http://arxiv.org/abs/hep-ph/0702100}{{\tt arXiv:hep-ph/0702100}}.
\bibitem[{Kurimoto et~al.(2002)Kurimoto, Li, and Sanda}]{Kurimoto:2001zj}
\bibinfo{author}{T.~Kurimoto}, \bibinfo{author}{H.-n. Li},
  \bibinfo{author}{A.~I. Sanda},
\newblock \bibinfo{title}{{Leading power contributions to $B \to \pi, \rho$
  transition form-factors}},
\newblock \bibinfo{journal}{Phys. Rev. D} \bibinfo{volume}{65}
  (\bibinfo{year}{2002}) \bibinfo{pages}{014007}.
  \DOIprefix\doi{10.1103/PhysRevD.65.014007}.
  \href{http://arxiv.org/abs/hep-ph/0105003}{{\tt arXiv:hep-ph/0105003}}.
\bibitem[{Hua et~al.(2021)Hua, Li, Lu, Wang, and Xing}]{Hua:2020usv}
\bibinfo{author}{J.~Hua}, \bibinfo{author}{H.-n. Li}, \bibinfo{author}{C.-D.
  Lu}, \bibinfo{author}{W.~Wang}, \bibinfo{author}{Z.-P. Xing},
\newblock \bibinfo{title}{{Global analysis of hadronic two-body B decays in the
  perturbative QCD approach}},
\newblock \bibinfo{journal}{Phys. Rev. D} \bibinfo{volume}{104}
  (\bibinfo{year}{2021}) \bibinfo{pages}{016025}.
  \DOIprefix\doi{10.1103/PhysRevD.104.016025}.
  \href{http://arxiv.org/abs/2012.15074}{{\tt arXiv:2012.15074}}.
\bibitem[{Liu et~al.(2008)Liu, Zhou, and Xiao}]{Liu:2008rz}
\bibinfo{author}{J.~Liu}, \bibinfo{author}{R.~Zhou}, \bibinfo{author}{Z.-J.
  Xiao},
\newblock \bibinfo{title}{{$B_s \to PP$ decays and the NLO contributions in the
  pQCD Approach}}  (\bibinfo{year}{2008}).
  \href{http://arxiv.org/abs/0812.2312}{{\tt arXiv:0812.2312}}.

\end{thebibliography}


\end{document}